\documentclass[aps,amsmath,amssymb,prd,showpacs,floatfix,preprint,superscriptaddress,nofootinbib,12pt]{article}
\usepackage{jheppub}
\pdfoutput=1

\usepackage{amsmath,epsfig}
\usepackage{amssymb,amsfonts}
\usepackage{latexsym}
\usepackage[latin1]{inputenc}
\usepackage{subeqnarray}
\usepackage{xcolor}

\usepackage{graphicx}

\relax
\def\be{\begin{equation}}
\def\ee{\end{equation}}
\def\bea{\begin{eqnarray}}
\def\eea{\end{eqnarray}}

\newcommand\fverb{\setbox\pippobox=\hbox\bgroup\verb}
\newcommand\fverbdo{\egroup\medskip\noindent%
                        \fbox{\unhbox\pippobox}\ }
\newcommand\fverbit{\egroup\item[\fbox{\unhbox\pippobox}]}

\newcommand{\bear}{\begin{eqnarray}}

\newcommand{\eear}{\end{eqnarray}}

\newcommand{\bsea}{\begin{subeqnarray}}
\newcommand{\esea}{\end{subeqnarray}}
\newbox\pippobox

\def\6{\partial}

\def\pa{\partial}

\def\m{\mu}

\def\sp{\;\;\;,\;\;\;}

\def\kk{{\mathcal{K}}}

\newcommand{\comments}[1]{}
\allowdisplaybreaks[3]

\renewcommand*{\thefootnote}{\fnsymbol{footnote}}
\title{Higher derivative corrections to incoherent metallic transport in holography}
\author[1,5]{Matteo Baggioli,}
\affiliation[1]{Institut de F\'isica d'Altes Energies (IFAE), Universitat Aut\'onoma de Barcelona, The Barcelona Institute of Science and Technology, Campus UAB, 08193 Bellaterra (Barcelona)}
\author[2,3,4]{Blaise Gout\'eraux,}
\affiliation[2]{Nordita, KTH Royal Institute of Technology and Stockholm University, Roslagstullsbacken 23, SE-106 91 Stockholm, Sweden}
\affiliation[3]{Stanford Institute for Theoretical Physics, Department of Physics, Stanford University, Stanford, CA 94305-4060, USA}
\affiliation[4]{APC, Universit\'e Paris 7, CNRS/IN2P3, CEA/IRFU, Obs. de Paris, Sorbonne Paris Cit\'e, B\^atiment Condorcet, F-75205, Paris Cedex 13, France (UMR du CNRS 7164).}
\author[4,5,6]{Elias Kiritsis\footnote{\href{http://hep.physics.uoc.gr/~kiritsis/}{http://hep.physics.uoc.gr/~kiritsis/}}}
\affiliation[5]{Crete Center for Theoretical Physics and I.P.P.,
Department of Physics, University of Crete, 71003 Heraklion, Greece}
\affiliation[6]{Crete Center for Quantum Complexity and Nanotechnology, University of
Crete, 71003 Heraklion, Greece}
\affiliation[7]{Institute of Theoretical Physics, School of Physics and Optoelectronic Technology, Dalian University of Technology, Dalian 116024, China}
\author[7,5]{and~Wei-Jia Li}
\emailAdd{mbaggioli@ifae.es}
\emailAdd{blaise.gouteraux@su.se}
\emailAdd{weijiali@dlut.edu.cn}

\abstract{Transport in strongly-disordered, metallic systems is governed by diffusive processes. Based on quantum mechanics, it has been conjectured that these diffusivities obey a lower bound $D/v^2\gtrsim \hbar/k_B T$, the saturation of which provides a mechanism for the T-linear resistivity of bad metals. This bound features a characteristic velocity $v$, which was later argued to be the butterfly velocity $v_B$, based on holographic models of transport. This establishes a link between incoherent metallic transport, quantum chaos and Planckian timescales. Here we study higher derivative corrections to an effective holographic action of homogeneous disorder. The higher derivative terms involve only the charge and translation symmetry breaking sector. We show that they have a strong impact on the bound on charge diffusion $D_c/v_B^2\gtrsim \hbar/k_B T$, by potentially making the coefficient of its right-hand side arbitrarily small. On the other hand, the bound on energy diffusion is not affected.}

\begin{document}
\subheader{CCTP-2016-20, CCQCN-2016-179, NORDITA-2016-129, SU-ITP-1621}
\maketitle
\renewcommand*{\thefootnote}{\arabic{footnote}}	
\setcounter{footnote}{0}
\section{Introduction}

It has long been argued that strongly-coupled quantum matter without quasiparticles has the shortest equilibration timescale allowed by quantum mechanics, $\tau_P\sim\hbar/k_B T$ \cite{PhysRevB.56.8714,sachdevbook, zaanen2004superconductivity}. This is believed to underpin many of the unusual transport properties of bad metals, like the $T$-linearity of their resistivity \cite{Bruin804,Hartnoll:2014lpa}, the violation of the Mott-Ioffe-Regel (MIR) bound \cite{HusseyMIR} or thermal diffusion \cite{Zhang:2016ofh}.

 If quasiparticles are short-lived, the dynamics is governed by the collective excitations of the strongly-coupled quantum fluid, which are simply the conserved quantities of the system (assuming no symmetry is spontaneously broken).

From the point of view of transport at late times, there are two distinct regimes, depending on the strength of momentum relaxation. When momentum relaxes slowly, thermoelectric transport is dominated by a single purely imaginary pole in the complex frequency plane, lying parametrically closer to the real axis than other `UV' poles. The dynamics is effectively truncated to keeping track only of this Drude-like pole, and the DC and AC electric conductivities take a simple form at low frequencies:

\begin{equation}
\label{ElCondMM}
\sigma(\omega)=\frac{\chi_{JP}^2}{\chi_{PP}(\Gamma-i\omega)}+O(\Gamma^0,\omega^0)\,,\qquad \sigma_{DC}=\frac{\chi_{JP}^2}{\chi_{PP}}\,.
 \end{equation}
The $\chi$'s are static susceptibilities and similar expressions hold for the other thermoelectric conductivities. $\Gamma$ is the momentum relaxation rate, and can be computed using the memory matrix formalism \cite{forster1990hydrodynamic,Hartnoll:2012rj,Mahajan:2013cja,Lucas:2015pxa} or gauge/gravity duality techniques \cite{Hartnoll:2012rj,Davison:2013jba,Davison:2013txa,Davison:2014lua,Davison:2015bea} by considering the operator breaking translation symmetry in the state. By assumption, $\Gamma\ll k_B T$ to avoid mixing with other, UV poles at scales $\sim k_B T$. DC conductivities in this regime are typically high and do not violate the MIR bound.

However, the optical conductivity of bad metals displays broad Drude peaks, with a width $\Gamma\sim1/\tau_P\sim T$, \cite{Bruin804}. This is the incoherent limit where momentum relaxes quickly and does not govern the late time transport properties. The collective excitations are simply diffusion of charge and energy \cite{Hartnoll:2014lpa}, as can be checked in explicit holographic models of incoherent transport \cite{Davison:2014lua}. In this case, DC conductivities are expected to be small, as there is no low-lying pole (compared to the temperature scale): this suggests an avenue towards violating the MIR bound, at least in principle.

Hartnoll conjectured \cite{Hartnoll:2014lpa} that the diffusivities obeyed a lower bound in this regime:
\begin{equation}
\label{DiffusionBound}
\frac{D_{e,c}}{v^2}\gtrsim\frac{\hbar}{k_B T}
\end{equation}
Here $v$ stands for some characteristic velocity of the system, which in  a weakly-coupled metal would be the Fermi velocity.
By making use of Einstein relations $D_c=\sigma/\chi$ (neglecting thermoelectric effects), a linear in $T$ resistivity follows when the bound is saturated, provided the charge static susceptibility carries no temperature dependence. Two questions come to mind when considering \eqref{DiffusionBound}: What is $v$ at strong coupling? Can the validity of this bound be tested in explicit models of incoherent transport?

Motivated by Gauge/Gravity duality computations, Blake proposed to replace $v$ in \eqref{DiffusionBound} by the ``butterfly velocity" $v_B$ \cite{Blake:2016wvh, Blake:2016sud}. Indeed, the butterfly velocity appears in certain out-of-time-order four-point correlation functions and is a measure of how fast quantum information scrambles. This provides a natural velocity at strong coupling, in contrast to the Fermi velocity which strictly speaking can only be defined in the presence of long-lived quasiparticles.

The butterfly velocity can be computed holographically in terms of horizon data by considering shockwave geometries \cite{Shenker:2013pqa,Blake:2016wvh,Roberts:2016wdl}, which encode the propagation of energy after a particle falls in the black hole horizon. The butterfly velocity is closely linked to the Lyapunov time $\tau_L$, which also obeys a lower bound featuring the Planckian timescale, $\tau_L\geqslant \hbar/2\pi k_B T$ \cite{Maldacena:2015waa}. This bound is saturated by quantum field theories with Einstein holographic duals. Thus, relating quantum chaos to incoherent metallic transport via Planckian timescales is  an appealing proposal.

Another hint comes from recent progress in computing holographic DC thermoelectric conductivities. It has been shown that these are given by formul\ae\ evaluated on the black hole horizon under very general assumptions \cite{Iqbal:2008by,Blake:2013bqa,Donos:2014uba,Donos:2014cya,Donos:2014yya,Donos:2015gia}. As the metric and matter field expansion close to the horizon are independent from details of the UV asymptotics, these formul\ae\ are in this sense universal. By way of the Einstein relations, the diffusivities are therefore connected to physics at the black hole horizon, as is the butterfly velocity.

\cite{Blake:2016wvh,Blake:2016sud} showed that the bound \eqref{DiffusionBound} held at low temperatures for particle-hole symmetric states which violate hyperscaling, both for exactly translation invariant black holes \cite{Charmousis:2010zz,Gouteraux:2011ce,Huijse:2011ef} as well as in the incoherent limit \cite{Donos:2014uba,Gouteraux:2014hca}. In these specific examples, the precise coefficient on the right-hand side of \eqref{DiffusionBound} is given in terms of the set of critical exponents, but is not expected to be universal. Of course, these holographic examples do not directly apply to bad metals, which are at finite density and not particle-hole symmetric. They do provide evidence that some version of the bound of \cite{Hartnoll:2014lpa} is at work when transport is diffusion-dominated. It is also important to note that no general proof of the bound \eqref{DiffusionBound} exists, as static susceptibilities depend in general on the full bulk solution and not just the horizon. Said otherwise, the diffusivities are not given by horizon formul\ae\ (though see the recent preprint where such a case is studied \cite{Blake:2016jnn}). More evidence for the bound \eqref{DiffusionBound} on energy diffusion was provided for finite density, AdS$_2$ horizons in \cite{Blake:2016jnn, Davison:2016ngz}.

In this work, our goal is to study the sensitivity of the combined proposal of \cite{Hartnoll:2014lpa,Blake:2016wvh, Blake:2016sud} to higher derivative terms in the effective holographic action. As the Einstein-Hilbert action is really only a leading two-derivative term in what should be thought of as a low energy effective action, it is natural to include higher-derivative terms. In passing, it also allows us to study the bound for a different class of finite density AdS$_2$ horizons than those of \cite{Blake:2016jnn, Davison:2016ngz}.

Holographic bounds and higher derivative corrections have a rich common history \cite{Kovtun:2004de,Brigante:2008gz,Grozdanov:2015qia,Grozdanov:2015djs,Baggioli:2016oqk,Gouteraux:2016wxj}. Whenever a bound of the kind \eqref{DiffusionBound} is formulated, the coefficient on the right-hand side of the inequality should really be understood as an $O(1)$ number:
\begin{equation}
\label{DiffusionBoundButterfly}
\frac{D_{c}}{v_B^2}\geq \mathcal A\frac{\hbar}{k_B T},\qquad\frac{D_{e}}{v_B^2}\geq \mathcal B\frac{\hbar}{k_B T},\qquad \mathcal A,\mathcal B\sim O(1)
\end{equation}
The name of the game is now to find out how higher-derivative terms affect $\mathcal A$ and $\mathcal B$, taking into account that: the higher-derivative couplings need to be small in some sense for the effective field theory approach to be well-defined; their allowed values are constrained by requiring the dual field theory to be causal. For instance, the KSS bound \cite{Kovtun:2004de} is lowered at most to
$$
{\eta\over s}\geq {16\over 25\pi}~ {\hbar\over k_B}
$$
 upon including a Gauss-Bonnet term \cite{Brigante:2008gz}, so that some version of the original bound is still believed to hold. On the other hand, while \cite{Grozdanov:2015qia} proved a lower bound on the electric conductivity in Einstein-Maxwell theory, in \cite{Baggioli:2016oqk,Gouteraux:2016wxj} it was shown how certain higher-derivative terms may lower this bound all the way to zero. That is to say, these couplings are sufficiently unconstrained by the stability analysis to allow in principle the coefficient on the right hand side of the bound to vanish.

The specific holographic models we will use to study the bound \eqref{DiffusionBoundButterfly} are given below in \ref{model1sec} and \ref{model2sec}.
They include quartic derivative terms between the Maxwell field strength and  the translation-symmetry breaking scalar sector.
The first contains the higher-derivative coupling $\frac{\mathcal{J}}{4}\,Tr[\mathcal{X}\,F^2]$ while the second contains  $\mathcal{K}\,Tr[\mathcal{X}]\frac{F^2}{4}$ where $\mathcal{X}$ involves the massless scalars and is defined in (\ref{X}).

Our main results is that while the bound on the diffusion of energy remains impervious to these terms, they strongly affect the diffusion of charge in the incoherent limit. For our two models, we find that
\begin{equation}
\frac{D_c\,T}{v_B^2}\geq\left(1-\frac32\mathcal J\right)\,\frac{1}{\pi}\,\frac{\hbar}{k_B}\,,\qquad \frac{D_c\,T}{v_B^2}\geq\left(1+6\,\mathcal K\right)\,f(\mathcal K)\,\frac{1}{\pi}\,\frac{\hbar}{k_B}
\end{equation}
with $f(\mathcal K)$ some function defined from \eqref{DiffBoundK}. Our analysis of stability and causality constraints restricts the couplings to
\begin{equation}
0\leq \mathcal J\leq2/3\,,\qquad -1/6\leq\mathcal K\leq1/6\,.
\end{equation}
Unlike higher-derivative corrections to the KSS bound, they seem to allow for an arbitrary violation of the bound \eqref{DiffusionBoundButterfly}, namely the right hand side may be tuned as small as desired. We pause here to note that it was already pointed out in \cite{Blake:2016sud} that the number $\mathcal B$ on the right hand side of the energy diffusion bound could be arbitrarily small, provided the dynamical exponent $z$ is also small. Inhomogeneous setups, both holographic or generalizations of SYK, also lead to violations of the bound featuring the butterfly velocity, \cite{Lucas:2016yfl,Gu:2017ohj}. There, it is shown that the inequality sign in \eqref{DiffusionBoundButterfly} is actually reversed.

Using higher derivative (gravitational) theories in order to investigate holographic phenomena is not without pitfalls. The actions we use in this paper do not lead to higher order than second derivatives in the classical equations of motion, and so do not contain ghosts. In the context of effective field theories, the higher-derivative couplings (including ours) should be considered as suppressed by appropriate powers of the Planck length or the effective string scale, and so do not typically give rise to causality violations in the absence of ghosts.

However, as pointed out above, we are also interested in situations where these corrections might be $\mathcal O(1)$. This happens for instance in classical (large $N$), weakly-coupled string theory: the curvature corrections to the Einstein-Hilbert action are set by the string coupling $\alpha'$, and become important at energies much lower than the Planck scale. Then \cite{Camanho:2014apa} showed that theories with such higher-derivative gravitational terms would necessarily violate causality, unless an infinite number of spin $\geq2$ particles were added at these energies. Their calculation amounts to showing that the higher derivative corrections can induce time advances in high energy scattering experiments in shockwave backgrounds, which in turn can lead to close timelike curves. We do not believe such causality violations can be triggered by the higher derivative terms we consider, since they do not involve higher derivatives of the metrics, which can be seen in \cite{Camanho:2014apa} to ultimately be the source of the time advances.
To summarize this discussion:

\begin{itemize}

\item Rigorously speaking we cannot fully trust truncated derivative corrections in string theory.

\item Experience from many  exact results in $\alpha'$ in string theory suggests that if the truncations do not violate basic principles of the theory (unitarity, the proper Cauchy problem etc), they  are expected to give qualitatively trustworthy results.

\item  We {\em do not} consider terms due to string loop corrections that may violate the large-N expansions at finite string coupling.
\end{itemize}    

We are therefore confident that the physics we analyze is characteristic of healthy higher-derivative corrections in string theory, and that our results give a glimpse into the finite coupling constant regime of the associated dual theories.

In the remainder of the paper, we present our results in more detail. Section \ref{sec1} is devoted to our holographic models, their black hole solutions and constraints coming from stability. In section \ref{sec2}, we present the expressions for their DC thermoelectric conductivities. In section \ref{sec3} we compute the charge and energy diffusion constants in the incoherent regime, and show how the charge diffusivity bound is affected by the higher derivative couplings. Some technical details are relegated to a number of appendices.

\section{The holographic models}\label{sec1}

Our starting point is the Einstein-Hilbert action in 4 bulk dimensions with negative cosmological constant $\Lambda$ (and $1/16\,\pi\, G_N=1$):
\begin{equation}
\mathcal{S}_g\,=\,\int\,d^4x\,\sqrt{-g}\,\left(\,R\,-\,2\,\Lambda\,\right)
\end{equation}
To accommodate finite density states, we add a U(1) vector field $A_\mu$ with associated field strength defined as $F_{\mu\nu}=\partial_{[\mu}A_{\mu]}$.
We will break translation invariance by introducing two massless St\"uckelberg fields with a bulk profile $\phi^I=k\,\delta^I_i\, x^i$ \cite{Andrade:2013gsa}. We construct the mixed tensor:
\begin{eqnarray}
{\mathcal{X}^\mu}_\nu\equiv \frac{1}{2}\sum_{I=x,y}\partial^\mu \phi^I \partial_\nu \phi^I={1\over 2}\sum_{I=x,y}g^{\m\rho}\pa_{\rho}\phi^I\partial_\nu \phi^I\,\,.
\label{X}\end{eqnarray}
and consider the generic action, coupling the electromagnetic and translation-symmetry breaking sectors:
\begin{eqnarray}
\mathcal{S}&=&\mathcal{S}_g\,+\,\mathcal{S}_a\\
\mathcal{S}_a&=&-\,\int\,d^4x\,\sqrt{-g}\,\mathcal{Z}\left(Tr[\mathcal{X}^m],\,Tr[\mathcal{X}^n\,F^2]\,\right)\,.
\end{eqnarray}
where
\be
Tr[\mathcal{X}^m]\,\equiv\, \mathcal{X}^\mu_{\,\,\,\,\nu_1\dots}\mathcal{X}^{\nu_{m-1}}_{\hspace{0.7cm}\mu}\sp
Tr[\mathcal{X}^n\,F^2]\,\equiv\,[\mathcal{X}^n]^\mu_{\,\,\,\,\nu}~F^{\nu}_{\hspace{0.2cm}\nu'}~F^{\nu'}_{\hspace{0.2cm}\mu}
 \ee
 and the indices run over non-negative integers $m,n\,=\,0,\,1,\,2\dots$.
  For convenience we also define $Tr[\mathcal{X}]\equiv X$.\\\\

We focus on the two following classes of  models:
\begin{itemize}
\item \textbf{Model 1:}
\begin{equation}
\label{Model1}
\mathcal{S}_a\,=\,-\,\,\int\,d^4x\,\sqrt{-g}\,\left(\,X\,+\,\frac{1}{4}\,F^2\,+\,\frac{\mathcal{J}}{4}\,Tr[\mathcal{X}\,F^2]\,\right)
\end{equation}
This model was introduced and analyzed recently in \cite{Gouteraux:2016wxj}.

\item \textbf{Model 2:}
\begin{equation}
\label{Model2}
\mathcal{S}_a\,=\,-\,\int\,d^4x\,\sqrt{-g}\,\,\mathcal{W}\left(X,F^2/4\right)
\end{equation}
This is a rather general class of models.  Within this class we will mostly focus on a special benchmark case:
\begin{equation}
\label{Benchmark2}
 \textbf{Model\,$2_U$}\,:\qquad\mathcal{W}(X,F^2/4)=X+U(X)\,\frac{F^2}{4}
\end{equation}
Moreover, in some cases we will specialize further and define:
\begin{equation}
\label{specificmod2}
 \textbf{Model\,$2_\mathcal{K}$}\,:\qquad U(X)\,=\,1\,+\,\mathcal{K}\,X
\end{equation}
also studied in \cite{Gouteraux:2016wxj}.
\end{itemize}
Furthermore we consider an isotropic ansatz for the bulk metric and other fields:
\begin{eqnarray}\label{bansatz}
ds^2=-D(r)\,dt^2+B(r)\,dr^2+C(r)\, dx^idx_i, \ \ A_\mu=A_t (r)\,dt, \ \ \phi^I=k\,\delta^I_i\, x^i,
\end{eqnarray}
where $i=x,y$ denotes the two spatial directions.

The aim of this paper is to study the effects of the the higher derivative terms \eqref{Model1}, \eqref{Model2} on the transport properties of the dual CFT at finite temperature T and charge density $\rho$.  If we set $\mathcal{J}=0$ or $U(X)=1$, then we recover the ``linear axion model''  of \cite{Andrade:2013gsa}.

\subsection{Model 1: the $\mathcal{J}$ coupling}\label{model1sec}

The $\mathcal{J}$ coupling does not affect the solution to the background equations given our Ansatz \eqref{bansatz}. This follows from how indices are contracted in $Tr(\mathcal{X}^n F^2)$ and it holds for all $n>1$. The background is then identical to the one found in \cite{Bardoux:2012aw,Andrade:2013gsa}:
\begin{align}
&ds^2\,=\,-\,D(r)\,dt^2\,+\,\frac{dr^2}{D(r)}\,+\,r^2\,dx^i\,dx_i\,,\nonumber\\
&D(r)\,=\,r^2\,\left[\,1\,-\,\frac{r_h^3}{r^3}\,-\,\left(\frac{k^2}{2\,r^2}\,+\,\frac{\mu^2\,r_h}{4\,r^3}\right)\,\left(1\,-\,\frac{r_h}{r}\right)\,\right],\nonumber\\
&A\,\equiv A_tdt=\,\left(\mu-\frac{\rho}{r}\right)\,dt
\end{align}
where we fix $\Lambda=-3$, and $r_h$ is the location of the event horizon.

Regularity of the gauge field at the horizon implies that we have $\rho=\mu\,r_h$, and the temperature of the background can be identified with the surface gravity at the horizon:
\begin{equation}
T\,=\,\frac{D'(r_h)}{4\,\pi}\,=\,\frac{3 \,r_h}{4 \,\pi }\,-\frac{k^2}{8 \,\pi  \,r_h}-\frac{\mu ^2}{16\, \pi  \,r_h}
\end{equation}

These are the background data we will use later in computing the conductivities.

\subsection{Model 2: $\mathcal{W}(X,F^2/4)$ action}\label{model2sec}

This class of models represents a generalization of what was already presented and studied in \cite{Gouteraux:2016wxj,Baggioli:2014roa,Alberte:2015isw,Baggioli:2016oqk,Baggioli:2016oju}. To simplify notation, we define
\begin{equation}
Y={1\over 4}F^2 \sp
\mathcal{W}_{Y}(Y,X)\equiv\frac{\partial \mathcal{W}(Y,X)}{\partial Y}\sp \mathcal{W}_{X}(Y,X)\equiv\frac{\partial \mathcal{W}(Y,X)}{\partial X}
\end{equation}

The solution for the background metric takes the form:
\begin{align}
&ds^2\,=\,-\,D(r)\,dt^2\,+\,\frac{1}{D(r)}\,dr^2\,+\,r^2\,dx^i\,dx_i\,,\nonumber\\
&D(r)\,=\,\frac{1}{2\,r}\int_{r_h}^r d\tilde{r} \left[6\,\tilde{r} ^2-\mathcal{W}(\bar{Y},\bar{X})\,\tilde{r} ^2-\frac{\rho^2}{\tilde{r} ^2\,\mathcal{W}_{Y}(\bar{Y},\bar{X})}\right]\,,
\end{align}
where $r=r_h$ is again the position of the event horizon.
The time component of the Maxwell equations for the gauge field $A=A_t(r)dt$ yields:
\begin{equation}
\rho=r^2\,\mathcal{W}_{Y} \left(\bar{Y},\bar{X}\right)\,A_t'
\end{equation}
where the constant $\rho$ represents the charge density of our system.
The background values $\bar X,\bar Y$ for the $X,Y$ scalar invariants turn out to be:
\begin{equation}
\bar{X}\,=\,\frac{k^2}{r^2}\,,\qquad\bar{Y}\,=\,\frac{1}{2}\,A_t'(r)^2
\end{equation}

The temperature of the solution is given as always by:
\begin{equation}
T\,=\,\frac{D'(r_h)}{4\,\pi}\,=\,\frac{1}{4\,\pi}\left[3\,r-\frac{\mathcal{W}(\bar{Y},\bar{X})\,r}{2}-\frac{\rho^2}{2\,r^3\,\mathcal{W}_{Y}(\bar{Y},\bar{X})}\right]_{r=r_h}
\end{equation}
More details about the specific models $2_U$ and $2_\mathcal{K}$ are presented in appendix \ref{appdetails}. In particular, when $\mathcal K<0$, some care must be exercised to derive the background solution. However, physical quantities expressed in terms of field theory data $(T,\mu)$ can safely be analytically continued to from $\mathcal K>0$ to $\mathcal K<0$.

\subsection{Stability}\label{secS}

The higher-derivative couplings $\mathcal J, \mathcal K$ were constrained in \cite{Gouteraux:2016wxj} by imposing positivity of the DC electric conductivity and studying the stability of the $a_x$ linear perturbation at zero density:
\begin{equation}
\label{StabilityConditions}
0\,\leq\,\mathcal{J}\,\leq 2/3\,,\qquad\qquad -1/6\,\leq\,\mathcal{K}\,\leq\,1/6\,.
\end{equation}
Here it is worth emphasizing that only the lower bound on $\mathcal{K}$ comes from considering the stability of the linear fluctuations at non-zero frequency -- a significantly harder problem than in the DC limit, where closed form expressions for all DC conductivity can be obtained and their inspection yields the other constraints.

We have extended the analysis in \cite{Gouteraux:2016wxj} by looking at both at background and linearized probes. The null energy condition (NEC) and the local thermodynamic stability (positivity of the specific heat and charge susceptibility) can be studied directly from the background solution. We find that the static susceptibilities are positive for all values of the higher-derivative couplings and do not constrain them at all. On the other hand, the NEC requires $\mathcal J\leq2/3$ and $\mathcal K\geq-1/6$. Further details are given in appendix \ref{AppCons}.  Here we simply comment on the NEC. It implies in general
\begin{equation}
\bar{X}\,\mathcal{W}_X(\bar{Y},\bar{X})\,-\,2\,\bar{Y}\,\mathcal{W}_Y(\bar{Y},\bar{X})\,\geq\,0\,.
\end{equation}
where $\bar{X},\bar{Y}$ are the background values for $X,Y$.\\
This constraint coincides with the absence of ghosts and matches with previous studies \cite{Baggioli:2014roa,Baggioli:2016oqk,Baggioli:2016oju,Gouteraux:2016wxj}. In particular it leads to a positive effective graviton mass squared $m_g^2\geq0$.

Extending to linear fluctuations, we could perform two checks: the stability of the parity-odd fluctuations at zero wavevector and zero density; and the analysis of the scaling dimensions of the IR operators in the AdS$_2\times R^2$ zero temperature spacetime, both in the transverse and longitudinal sector and at non-zero wavevector $q$. If these dimensions become complex for certain values of the couplings and a certain range of wavevectors, we have found an instability.

At non-zero density, the linear fluctuation equations are coupled and we could not rewrite them as decoupled Schr\"odinger equations. One way to confirm our stability analysis would be to inspect the spectrum of quasi-normal modes and check they are all in the lower half of the complex frequency plane. This analysis would be quite involved and beyond the scope of this paper. So we content ourselves with the necessary conditions \eqref{StabilityConditions}.

The analysis of the scaling dimensions of the IR operators is simplest when the linear equations around AdS$_2\times R^2$ can be decoupled in terms of gauge-invariant master variables. These decoupled equations can be integrated, imposing ingoing boundary conditions. And the scaling dimensions can be read off from the asymptotics of the resulting solutions. This program can only be carried out in very special, highly symmetric cases, like the AdS-Reissner-Nordstrom black hole. It does not seem possible in our setup, as the equations do not decouple. We can however work out the scaling dimensions by plugging in a power law Ansatz for the perturbations. Details of the derivation are provided in appendix \ref{AppDim}. We do not write here the final expressions for the scaling dimensions, which are very messy.

For the model 1, we could check analytically that the scaling dimensions are always real in the range $0\leq\mathcal J\leq2/3$. The model $2_{\mathcal K}$ is harder to analyze in full generality. We can show that the transverse scaling dimensions are real in the range $-1/6\leq {\mathcal K}\leq1/6$. In the longitudinal sector, we can only do this when picking random values for ${\mathcal K}$ in the same range  but we cannot prove it in general.

All in all, we take it that the arguments above make a very good case for stability of both models given the condition \eqref{StabilityConditions}.

\section{DC thermoelectric conductivities}\label{sec2}
Thermoelectric transport in the dual CFT can be described by the generalized Ohm's law:
\begin{eqnarray}\label{ohm}
\left(  \begin{array}{c}
  J^x \\
    Q^x \\
  \end{array}\right)=\left(
  \begin{array}{cc}
    \sigma & \alpha \,T \\
    \bar{\alpha}\, T & \bar{\kappa} \,T \\
  \end{array}
\right)
\left(  \begin{array}{c}
  E_x \\
    -\nabla_x T/T \\
  \end{array}\right)
\end{eqnarray}
where the matrix of thermoelectric conductivities parametrizes the linear response to electric fields and temperature gradients.
In the absence of parity violation, the conductivity matrix is symmetric $\alpha=\bar\alpha$.
DC conductivities can be computed holographically in terms of data on the black hole horizon using the techniques described in \cite{Donos:2014uba,Donos:2014cya}. We simply quote the final results in the main text and relegate the details of the computation to appendix \ref{app:DCcond}.

\subsection{Model 1}
The DC conductivities for the model described in \ref{model1sec} read as follows:
\begin{align}\label{DCmodel1}
&\sigma\,=\,1\,-\,\frac{\mathcal{J}\, k^2}{4 \,r_h^2}+\,\frac{\mu ^2 \left(1-\frac{\mathcal{J} \,k^2}{4 \,r_h^2}\right)^2}{k^2 \left(1+\frac{\mathcal{J} \,\mu ^2}{4\,r_h^2}\right)}\,,\\
&\bar{\kappa}\,=\,\frac{4\, \pi\,  s\, T}{k^2 \left(1+\frac{\mathcal J \,\mu ^2}{4\,r_h^2}\right)}\,,\\
&\alpha\,=\,\bar{\alpha}\,=\,\frac{4\,\pi\,\mu\,r_h\,\left(1-\frac{\mathcal{J} \,k^2}{4 \,r_h^2}\right)}{k^2 \left(1+\frac{\mathcal J \,\mu ^2}{4\,r_h^2}\right)}\,,\\
&\kappa\,=\,\frac{16\, \pi ^2\, r_h^2\, T}{k^2+\mu ^2}\,.
\end{align}
where the entropy density $s=4\,\pi\, r_h^2$.\\
Here, $\bar{\kappa}$ is the thermal conductivity at zero electric field, while $\kappa$ is the thermal
conductivity at zero current. They are related through $\kappa=\bar{\kappa}-\,\bar{\alpha}\,\alpha\, T/\sigma$.
\subsection{Model 2}
The DC conductivities for the model described in \ref{model2sec} read:
\begin{align}\label{DCmodel2}
&\sigma=\left[\mathcal{W}_{Y}\left(\bar{Y},\bar{X}\right)+\frac{4\,\pi\,\rho^2}{k^2\,s\,\mathcal{W}_{X}\left(\bar{Y},\bar{X}\right)}\right]_{r=r_h},  \\
&\alpha=\bar{\alpha}=\frac{4\,\pi\,\rho}{k^2\,\mathcal{W}_{X}\left(\bar{Y},\bar{X}\right)}\Big|_{r=r_h},\\
&\bar{\kappa}=\frac{4 \,\pi\, s\,T}{ k^2\,\mathcal{W}_{X}\left(\bar{Y},\bar{X}\right)}\Big|_{r=r_h}   \\
&\kappa=\frac{4\,\pi\, s\,T}{k^2\,\mathcal{W}_{X}\left(\bar{Y},\bar{X}\right)+\frac{4\,\pi\,\rho^2}{s\,\mathcal{W}_{Y}\left(\bar{Y},\bar{X}\right)}}\Big|_{r=r_h}.
\label{22}
\end{align}
We can additionally define the Lorentz ratios:
\begin{align}
&\bar{L}\,\equiv\,\frac{\bar{\kappa}\,T}{\sigma}\,=\,\frac{4\,\pi\, s^2\,T^2}{k^2\,s\,\mathcal{W}_Y\,\mathcal{W}_X\,+\,4\,\pi\, \rho^2}\,,\\
&L\,\equiv\,\frac{\kappa\,T}{\sigma}\,=\,\frac{4\,\pi\, k^2\, s^3\, \mathcal{W}_Y\,\mathcal{W}_X\,T^2}{\left[k^2\,s\,\mathcal{W}_Y\,\mathcal{W}_X\,+\,4\,\pi\, \rho^2\right]^2}\,.
\end{align}
The values of the DC transport coefficients (\ref{DCmodel2})-(\ref{22}) for the particular models $2_U$ and $2_\mathcal{K}$ are presented in appendix \ref{appdetails}.\\[0.5cm]

\paragraph{About the Kelvin formula\\}
Recently, the relation
\begin{equation}
\label{Kelvin}
\frac{\alpha}{\sigma}\Big|_{T=0}\,\equiv\,\lim_{T\rightarrow 0}\frac{\partial s}{\partial \rho}\Big|_T
\end{equation}
has been highlighted as a feature of any AdS$_2\times R^2$ horizon \cite{Blake:2016jnn,Davison:2016ngz}. \cite{Davison:2016ngz} argued further that this was fixed by the symmetries of AdS$_2$. Indeed, we observe that \eqref{Kelvin} is verified in all the models we considered. We give more details about this check in appendix \ref{AppKelvin}.

\section{Impact of higher derivative couplings on the diffusivity bounds}\label{sec3}

The incoherent limit, \textit{i.e.} the limit of strong momentum dissipation, is defined by:
\begin{equation}
T,\mu\ll k
\label{i1}\end{equation}
while keeping the dimensionless ratio $T/\mu$ finite. This is the regime where transport is governed by diffusive processes \cite{Hartnoll:2014lpa} rather than by slow momentum relaxation, as was checked in the linear axion model \cite{Davison:2014lua}.

In this limit, both the off-diagonal conductivities decay faster with $k$ than the diagonal ones (which are actually non-zero). Effectively, the charge and heat flows decouple \cite{Davison:2015bea} in spite of the fact that this is not a zero density limit. The same is true for the matrix of static susceptibilities.
Consequently, in the incoherent regime, the charge and energy diffusivities can be independently defined as:
\begin{equation}
T,\mu\ll k:\qquad\qquad D_c\,=\,\frac{\sigma}{\chi}\,,\qquad D_e\,=\,\frac{\kappa}{c_v}
\end{equation}
where
\be\chi={\partial \rho\over \partial \mu}\Big |_{T}\sp c_v=T \,{\partial s\over \partial T}\Big |_{\mu}\,.
\ee
$\chi$  is the charge susceptibility at constant temperature
and $c_v$ the specific heat of the system at constant chemical potential (which in this limit is the same as at constant charge density).

The butterfly velocity of the system, describing the spreading of quantum information in the dual QFT, has been already computed in \cite{Blake:2016wvh} for a generic background of the form \eqref{bansatz} and it turns out to be:
\begin{equation}
v_B^2\,=\,\frac{2\,\pi\,T}{C'(r_h)}
\end{equation}
Because we have chosen a radial gauge so that $C(r)=r^2$ we obtain the general expression:
\begin{equation}
v_B^2\,=\,\frac{\pi\,T}{r_h}
\end{equation}
The linear axion model was defined in the beginning of section \ref {model1sec} and  corresponds to setting the  higher derivative couplings $\mathcal{J}$ and $\mathcal{K}$ to zero. In this model and in the incoherent limit defined in (\ref{i1}),  both the charge and the energy diffusivities, appropriately normalized, are bounded from below as shown in \cite{Blake:2016wvh} (in passing generalizing the analysis there to finite density),
\be
\frac{D_c\,T}{v_B^2}\geq \frac{1}{\pi}\sp  \frac{D_e\,T}{v_B^2}\geq\frac{1}{2\,\pi}\;.
\ee

Our aim is to investigate if this inequality is still valid once higher derivative corrections are taken in consideration.
\vskip .5cm

\paragraph{Model 1: the $\mathcal{J}$ coupling\\}

Since the background is not affected by $\mathcal{J}$, it is straightforward to perform the same computations at finite $\mathcal{J}$. The susceptibility is given by:
\begin{equation}
\chi\,=\,r_h\;,
\label{a1}\end{equation}
while the conductivity in the incoherent limit is:
\begin{equation}
\sigma_{DC}^{(inc)}\,=\,\left(1\,-\,\frac{3}{2}\,\mathcal{J}\right)
\label{a2}\end{equation}
We note that in the incoherent limit, the radius of the horizon becomes proportional to the momentum dissipation strength $k$.
In particular, for models 1, $2_U$ and $2_{\kk}$ considered in this paper,  we have
\be
r_h\,=\, k/\sqrt{6}\;.
\label{a3} \ee
 The equation above implies that in such models the butterfly velocity in the incoherent limit becomes:
\begin{equation}
\left(v_B^2\right)^{(inc)}\,=\,\frac{\sqrt{6}\,\pi\,T}{k}
\label{a4}\end{equation}
In addition, the heat capacity and the thermal conductivity in the combined incoherent limit are given by:
\begin{equation}\label{endiff}
\kappa^{(inc)}\,=\,\frac{8\,\pi^2}{3}\,T,\qquad c_v^{(inc)}\,=\,\frac{8}{3}\,\sqrt{\frac{2}{3}}\,\pi^2\,T\,k\,.
\end{equation}
Using (\ref{a1})-(\ref{endiff}) we obtain the following equalities in this limit
\be \frac{D_c\,T}{v_B^2}\Big |_{inc}= \frac{1}{\pi}\,\left(1\,-\,\frac{3}{2}\,\mathcal{J}\right)\sp
\frac{D_e\,T}{v_B^2}\Big |_{inc}=\frac{1}{2\,\pi}\;.
\ee

The charge diffusivity is modified to leading order in the incoherent limit and the dimensionless ratio $\frac{D_c\,T}{v_B^2}$ vanishes  for $\mathcal{J}=2/3$. At that same value of $\mathcal{J}$ the incoherent DC conductivity $\sigma_{DC}^{(inc)}$ vanishes.

We believe this  to be a generic feature of all effective actions  where momentum relaxing terms  couple directly to the Maxwell term.
 We  obtain the same results considering higher order deformations of the type:
\begin{equation}
\sim\,\,Tr\left(\mathcal{X}^n\,F^2\right)\,,\qquad \text{with}\quad n>1\,.
\end{equation}
for all of which the background would still remain unchanged.

\vskip .5cm

\paragraph{Model $2_U$\\}

We will now investigate the $2_U$ class of models defined as
\begin{equation}
\mathcal{W}(X,Y)\,=\,X\,+\,U(X)\,Y\;.
\end{equation}

In this case, the static susceptibility in the incoherent limit is:
\begin{equation}
\chi\,=\,\left(\,\int_{r_h}^{\infty}\,\frac{1}{y^2\,U(k^2/y^2)}\,dy\,\right)^{-1}
\end{equation}
The precise derivation of this formula is shown in appendix \ref{appdetails}.
The susceptibility above is finite, because $U(0)=1$ (in order to have the correctly normalized Maxwell term near the boundary) but it is manifestly not given in terms of horizon data and depends on the full bulk geometry \cite{Iqbal:2008by,Blake:2016wvh}.

The DC conductivity in the incoherent limit can be extracted from the generic
formul\ae\ of the previous section and it reads:
\begin{equation}
\sigma_{DC}^{(inc)}=\,U(k^2/r_h^2)
\end{equation}
Combining the previous results we conclude that in the combined incoherent limit the diffusivity asymptotes to
\begin{equation}
\frac{D_c\,T}{v_B^2}\Big|_{inc}\,=\,\lim_{\substack{\\T\,\to\, 0 \\ \mu\,\to\, 0}}\,\frac{U(k^2/r_h^2)\,r_h}{\pi}\,\int^{\infty}_{r_h}\,\frac{1}{y^2\,U(k^2/y^2)}
\end{equation}
The key point is that the dimensionless ratio $\frac{D_c\,T}{v_B^2}$ becomes zero every time $U(X)$ vanishes in the incoherent limit. This is the same point were $\sigma_{DC}^{(inc)}$ vanishes. We find this correlation robust and present in all the models we considered.

\begin{figure}
\centering
\includegraphics[width=.4\textwidth]{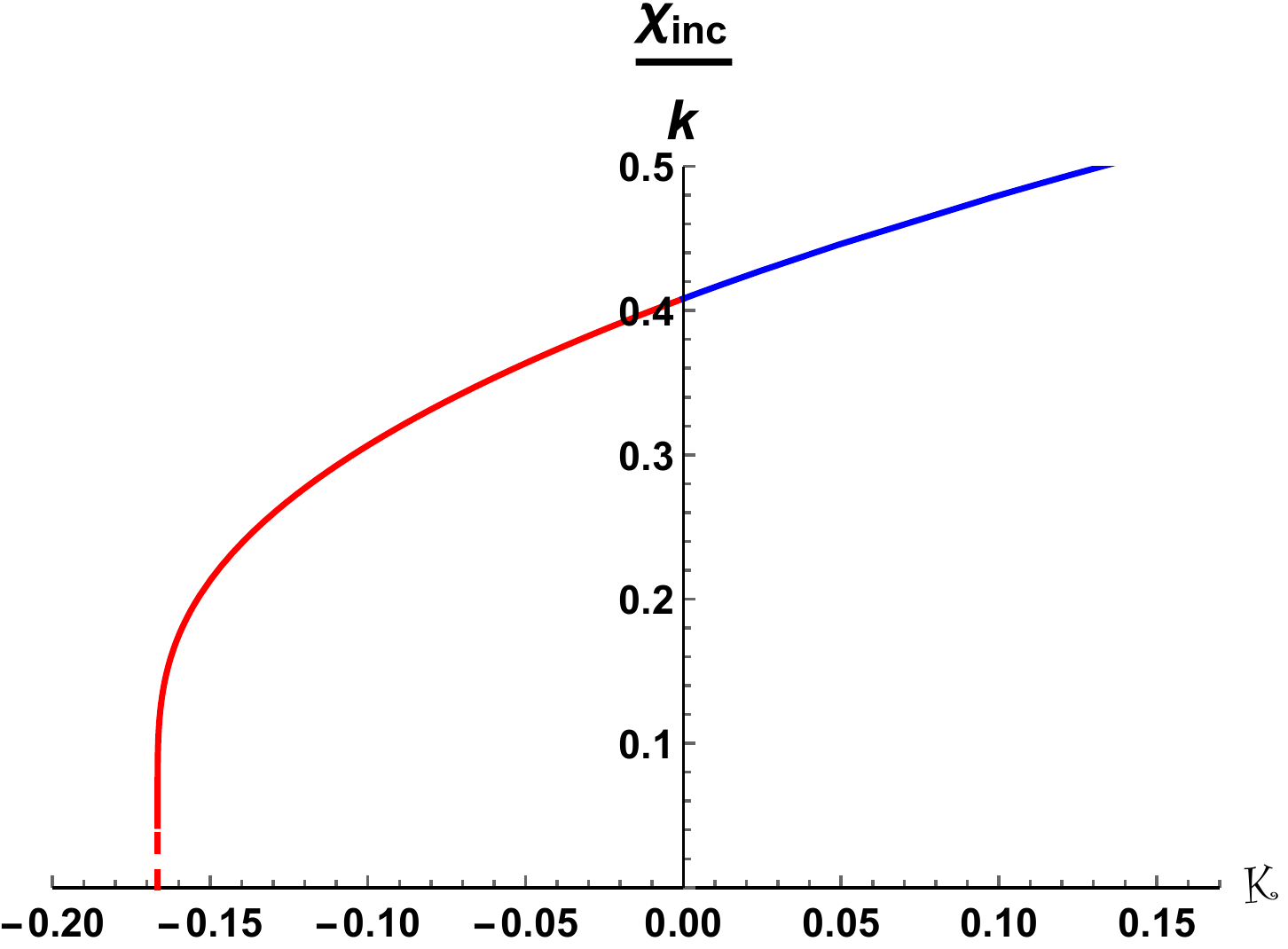}%
\qquad
\includegraphics[width=0.4\textwidth]{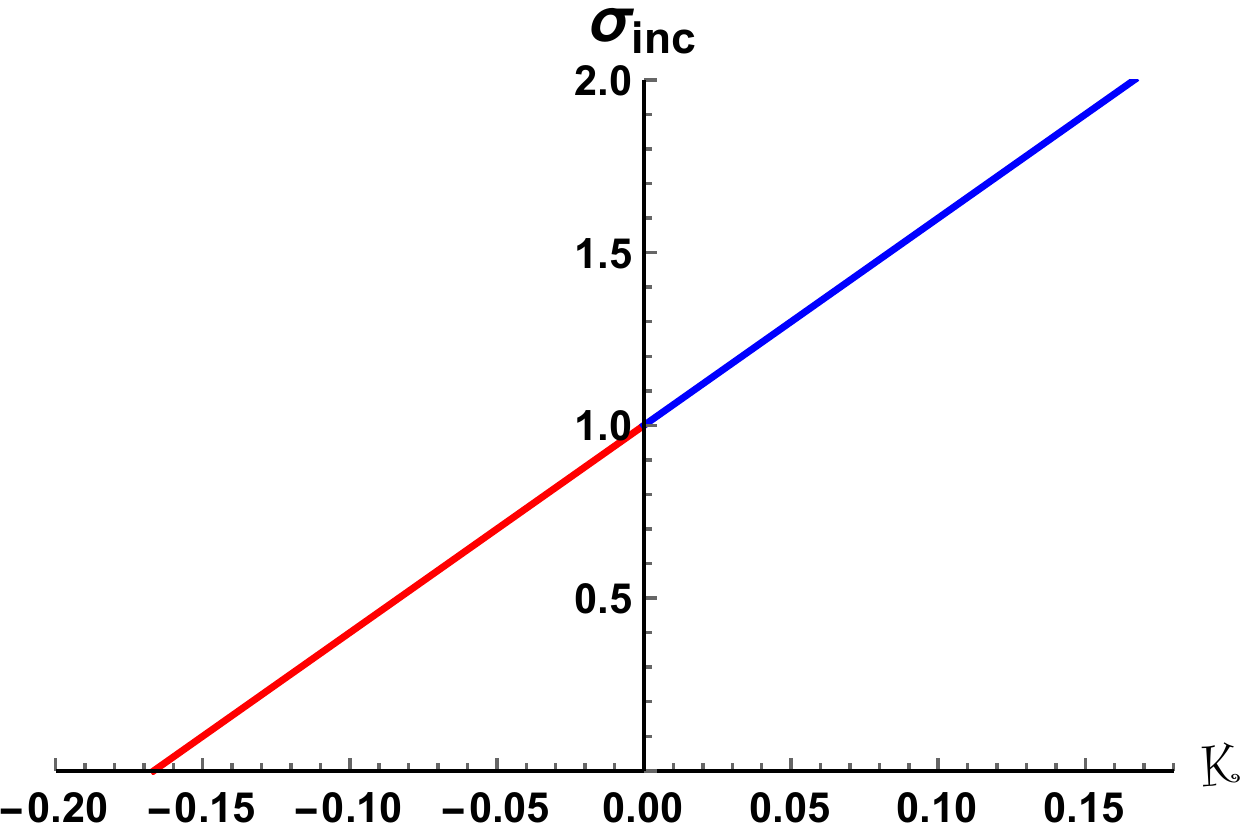}
\caption{Physical quantities in the incoherent limit for the specific model $2_\mathcal{K}$: $U(X)=1+\mathcal{K}X$. \textbf{Left: } Susceptibility in units of k as a function of $\kk$. \textit{The dashed red line has been added manually because Mathematica was not able to plot the function in its whole domain}.  \textbf{Right :} Incoherent conductivity as a function of $\kk$.}
\label{figQ}
\end{figure}
On the other hand we can show  that the $U(X)$ coupling does not affect energy diffusion and we still have:
\begin{equation}
\frac{D_c\,T}{v_B^2}\Big |_{inc}=\frac{1}{2\,\pi}
\end{equation}
This is due to the fact that the value of the heat capacity and the thermal conductivity in the combined incoherent limit are not modified by the U coupling and they still take the form indicated in (\ref{endiff}).

\vskip.5cm

\paragraph{Model $2_\mathcal{K}$: the $\mathcal{K}$ coupling\\}

To illustrate the previous paragraph, we choose the function:
\begin{equation}
U(X)\,=\,1\,+\mathcal{K}\,X
\end{equation}

In the allowed range of parameters, the DC conductivity in the incoherent limit is given by:
\begin{equation}\label{f2}
\sigma_{DC}^{(inc)}=1+6\,\mathcal{K} \quad \text{with } -1/6\leq\,\mathcal{K}\,\leq 1/6
\end{equation}
and the charge susceptibility by:
\begin{align}
\label{f1}
&\chi^{(inc)}=\begin{cases}
    \frac{2 \,k\, \sqrt{\mathcal{K}}}{\pi -2 \,ArcTan\left(\frac{1}{ \sqrt{6\,\mathcal{K}}}\right)}       & \quad \text{if }\,\,\,\, 0<\mathcal{K}\leq1/6\\[0.2cm]
    1/\sqrt{6} & \quad \text{if }\,\,\mathcal{K}\,=\,0\\[0.2cm]
    \frac{k\,  \sqrt{|\mathcal{K}|}}{\log \left(\frac{\sqrt{6\,|\mathcal{K}|}+1}{\sqrt{1-6 \,|\mathcal{K}|}}\right)}  & \quad \text{if } \,\,-1/6\leq\mathcal{K}<0\\
  \end{cases}
\end{align}
These two quantities are shown in fig.\ref{figQ}.
The incoherent heat capacity and the thermal conductivity are not affected by the $\mathcal{K}$ coupling and they take the form \eqref{endiff}.

Using the definition of the butterfly velocity given previously, we compute the diffusivities and obtain the dimensionless ratios:
\begin{align}
\label{DiffBoundK}
&\frac{D_c\,T}{v_B{}^2}\Big |_{inc}=\begin{cases}
    \frac{(6 \,\mathcal{K}+1) \left(\pi\,-\,2\, ArcTan\left(\frac{1}{\sqrt{6 \,\mathcal{K}} }\right)\right)}{2 \,\pi \, \sqrt{6\,\mathcal{K}}}\,+\,\dots       & \quad \text{if }\,\,\,\, 0<\mathcal{K}\leq1/6\\[0.2cm]
    \frac{1}{\pi}& \quad \text{if }\,\,\mathcal{K}\,=\,0\\[0.2cm]
    \frac{(1-6 \,|\mathcal{K}|) \log \left(\frac{ \sqrt{6\,|\mathcal{K}|}+1}{\sqrt{1-6 \,|\mathcal{K}|}}\right)}{\pi\, \sqrt{6\,|\mathcal{K}|}}\,+\,\dots  & \quad \text{if } \,\,-1/6\leq\mathcal{K}<0\\
  \end{cases}\\[0.2cm]
  &\frac{D_e\,T}{v_B{}^2}\Big |_{inc}=\frac{1}{2\,\pi}
\end{align}
\begin{figure}
\centering
\includegraphics[width=0.5\textwidth]{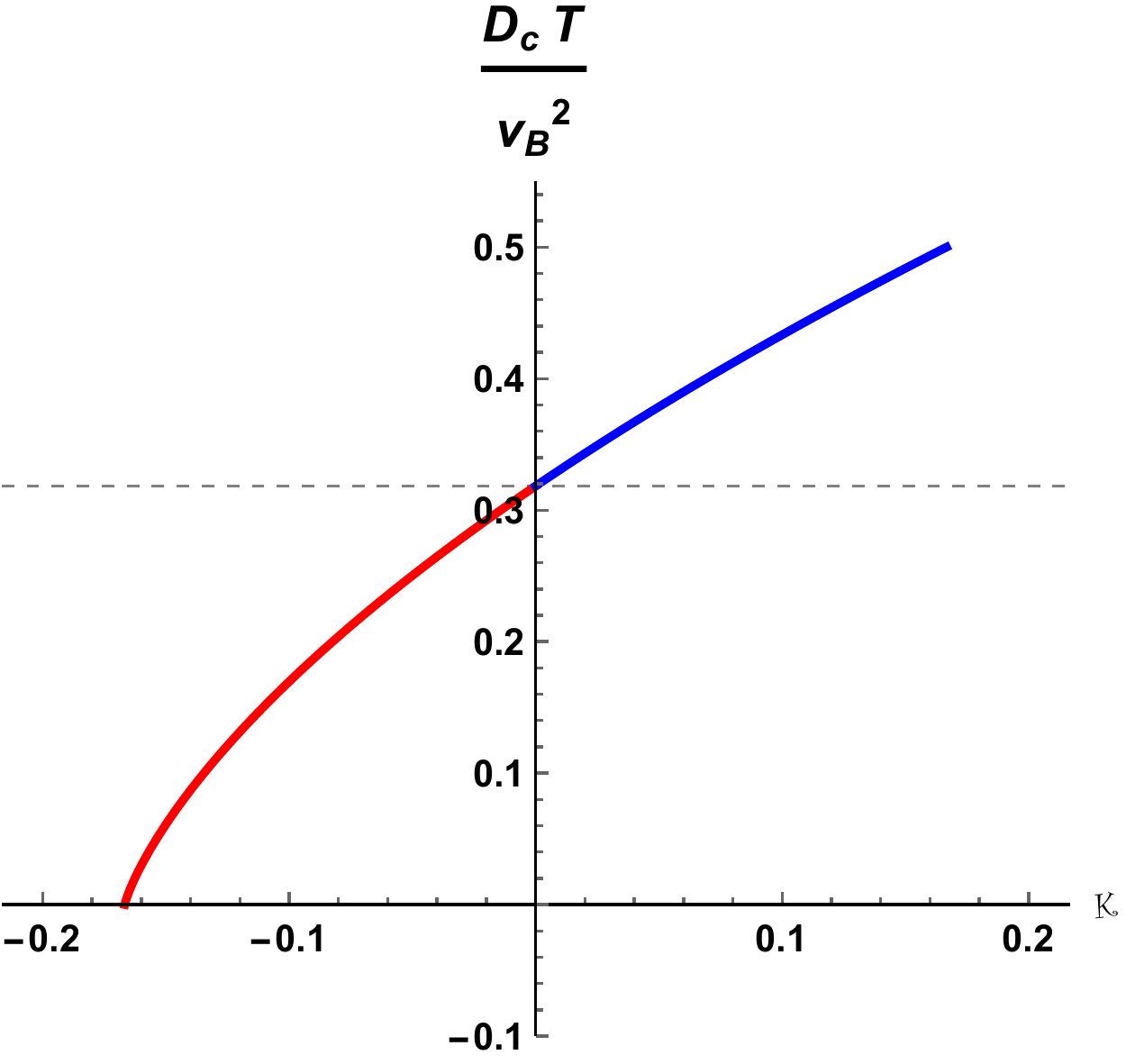}
\caption{Incoherent limit of $\frac{D_c\,T}{v_B^2}$ in function of the $\mathcal{K}$ parameter for the choice $U(X)=1+\mathcal{K}X$. The dashed line is the bound in the case $\mathcal{K}=0$.}
\label{figU}
\end{figure}
The behaviour of $D_c\,T/{v_B}^2$ in function of $\mathcal{K}$ is shown in fig.\ref{figU}. It vanishes when $\mathcal{K}=-1/6$, at the boundary of the stability region. There, the DC conductivity vanishes linearly while the charge susceptibility does so only logarithmically. The version of the charge diffusivity bound proposed in \cite{Blake:2016wvh, Blake:2016sud} can be violated in this model as well. 

In contrast, the higher-derivative term does not affect the ratio $D_e\,T/{v_B}^2$.
\vskip.5cm
\paragraph{Arbitrary St\"uckelberg potential $V(X)$\\}
As we have seen above, the bound on the diffusion of energy is not affected by the higher-derivative couplings we have turned on. A natural extension is to introduce an arbitrary potential $V(X)$ for the St\"uckelberg fields, rather than the linear version $V(X)=X$ we have been using throughout the draft. However, as we show below, this has no effect on the diffusion of energy in the incoherent limit. For simplicity, we consider $W(X,F^2)=V(X)$ in \eqref{Model2}, that is we consider zero density states.\\
The temperature of the model is defined by:
\begin{equation}
T\,=\,\frac{3 \,r_h}{4 \,\pi }-\frac{r_h\, V\left(\frac{k^2}{r_h^2}\right)}{8 \,\pi }
\end{equation}
The radius of the horizon in the incoherent limit is still proportional to the momentum dissipation strength k via the relation:
\begin{equation}\label{pp}
r_h^{(inc)}\,=\,\frac{k}{\sqrt{V^{-1}(6)}}
\end{equation}
which is in agreement with \eqref{a3} if we set $V(X)=X$.\\
In addition the thermal conductivity and the heat capacity are defined by:
\begin{equation}
\kappa\,=\,\frac{4\,\pi\,s\,T}{k^2\,V'\left(\frac{k^2}{r_h^2}\right)}\,,\qquad c_v\,=\,8\,\pi\,r_h\,T\,\left(\frac{dT}{d r_h}\right)^{-1}=\frac{64 \,\pi ^2 \,r_h^3\, T}{2\, k^2\, V'\left(\frac{k^2}{r_h^2}\right)-r_h^2 \left(V\left(\frac{k^2}{r_h^2}\right)-6\right)}
\end{equation}
In the incoherent limit \eqref{pp} we discover that their ratio reads as:
\begin{equation}
\frac{\kappa}{c_v}\Big|_{inc}\,=\,\frac{\sqrt{V^{-1}(6)}}{2\,k}
\end{equation}
Once we combine the previous result with the definition of the butterfly velocity we obtain:
\begin{equation}
\frac{D_e\,T}{v_B^2}\Big|_{inc}\,=\,\frac{\kappa}{c_v}\,\frac{T}{v_B^2}\Big|_{inc}\,=\,\frac{1}{2\,\pi}
\end{equation}
which is the same expression found for the linear choice $V(X)=X$.\\
Therefore we conclude that the $V(X)$ generalization has no impact on the energy diffusion and the following inequality
\begin{equation}\label{uno}
\frac{D_e}{v^2}\gtrsim\frac{\hbar}{k_B T}
\end{equation}
still holds.

We observe that this originates from two successive cancellations, such that in the end the general potential $V(X)$ does not affect the bound \eqref{uno}. Firstly, some factors of $V'(X)$ drop out when computing the energy diffusivity in the incoherent limit. Secondly, the remaining factor $V^{-1}(6)$ in \eqref{pp} is compensated by an analogous term in the expression for the butterfly velocity, leading finally to \eqref{uno}.

\section{Discussion}

In this paper, we studied higher derivative couplings $g_i$ between the charge and translation symmetry breaking sectors in toy-models of holographic thermoelectric transport. Focusing on the limit of fast momentum relaxation, we pointed out that these terms have a very strong impact on a recently proposed bound on charge diffusion \cite{Blake:2016wvh, Blake:2016sud} (elaborating on a previous proposal \cite{Hartnoll:2014lpa}):
\begin{equation}
\frac{D_c}{v_B^2}\gtrsim\frac{\hbar}{k_B T}
\end{equation}
where $v_B$ is the butterfly velocity. While the proposal in \cite{Hartnoll:2014lpa} essentially came from general considerations as well as experimental data on so-called bad metals, its refinement in \cite{Blake:2016wvh, Blake:2016sud} was justified using holographic computations. As such, it is rather natural to test it further by including higher derivative terms in the effective holographic action.

For simplicity, we restricted our investigation to models with quartic couplings only (see section \ref{sec1}). We paid particular attention to the stability and the consistency of the models restricting the allowed values for the couplings:
\begin{equation}
g_i^{min}\,\leq\,g_i\,\leq\,g_i^{max}
\end{equation}
where the edge values depend on the specific features of the model (see section \ref{secS}).\\

In more detail for all those cases we found a relation of the type:
\begin{equation}
\frac{D_c}{v_B^2}\,=\,\mathcal{A}\left(g_i\right)\,\frac{\hbar}{k_B T}
\end{equation}
where $\mathcal{A}$ is an order one number which only depends on the higher derivative couplings. It vanishes for particular finite values of the higher derivative couplings $g_i^{*}$. Of course one should keep in mind that these higher derivative couplings should be suppressed by powers of the string coupling, so it is unclear how realistic the values leading to $\mathcal{A}(g_i^{*})=0$ are. We note that since no higher derivative gravitational term is involved, the couplings we consider may be $\mathcal O(1)$ without violating causality along the lines of \cite{Camanho:2014apa}. It is very intriguing that the values $g_i^{*}$ lie at the edge of the range allowed by the stability analysis. This is also true for the hyperscaling violating metrics examined in \cite{Blake:2016wvh, Blake:2016sud}. It would be worthwhile to understand this better.\footnote{We are grateful to Mike Blake for discussions on this point.}\\

Let us pause to compare with the analogous violation of the KSS bound $\eta/s\geq \hbar/4\pi k_B$ by higher derivative terms, like Gauss-Bonnet \cite{Brigante:2007nu}. Including these terms modify the order one number on the right hand side and indeed can lower it, but causality prevents its vanishing. So the notion that there should be a lower bound on the ratio of shear viscosity to entropy density in strongly-coupled quantum field theories still survives. Our case is crucially different since the violation can be arbitrary down to zero value, at least up to the validity of our stability analysis. Admittedly, we have not fully carried it out as the lack of decoupling of the fluctuation equations render it untractable analytically. A more elaborate numerical analysis is needed and beyond the scope of this work. We hope to return to it in the future.\\

Two more features of our analysis, specifically due to the incoherent limit $T,\mu\ll k$ are very noteworthy. First, thermal and electrical transport always decouple. This was already noted in \cite{Davison:2015bea}: there, two decoupled, gauge-invariant bulk variables were found to be dual to two decoupled currents (with, in the language of \eqref{ElCondMM}, zero off-diagonal static susceptibility), which in the incoherent limit asymptoted to the charge and heat currents respectively. The same physical mechanism is at work here, upon turning on higher-derivative terms: while we have not been able to find decoupled bulk variables, the off diagonal elements of the conductivity and susceptibility matrices decay faster than the diagonal one. It would be very interesting if this was a general feature of thermoelectric incoherent transport, beyond these specific holographic examples.

We also found a strong correlation between the vanishing of the dimensionless parameter $\mathcal{A}$, which controls charge diffusion, and the vanishing of the corresponding DC electric conductivity in the incoherent limit. In all the models we considered the charge susceptibility remains finite in the incoherent limit implying the relation:
\begin{equation}
\mathcal{A}\left(g_i^{*}\right)\,=\,0\qquad\Longleftrightarrow\qquad \sigma_{DC}^{(inc)}\left(g_i^{*}\right)\,=\,0
\end{equation}
In other words, the charge diffusion bound is badly violated every time the corresponding incoherent electric DC conductivity vanishes. One way out would be if the bound shown in \cite{Grozdanov:2015qia} in four-dimensional Einstein-Maxwell theories could be generalized to our setup. But as we have argued, and unless a more refined stability analysis narrows the allowed range for the couplings, this does not seem to be the case.\\

A relevant question here is to what extend such a bound is independent  from those on the conductivity \cite{Baggioli:2016oqk,Gouteraux:2016wxj}. This depends on the behaviour of the static susceptibility and butterfly velocity in the incoherent limit.

In our model 1, the ratio $v_B^2\chi/T$ is $T$ and $r_h$ independent. This is a special feature of this model, whereby the background is not affected by the higher-derivative coupling. It affects only transport and so indeed the vanishing of the diffusivity bound follows from the vanishing of the dc conductivity.

The behaviour of the charge diffusivity is less trivial in the model with the $\mathcal K$ coupling, as seen from eqns \eqref{f1}-\eqref{DiffBoundK}. Background thermodynamics are affected by the higher-derivative coupling. From \eqref{f1}, we see that $\chi\to 0$ when $\mathcal K\to-1/6$. This means that in this limit, no electric current propagates (the dc conductivity is zero), and introducing a small chemical potential does not create a charge density in linear response (the susceptibility is zero).
However, the dc conductivity vanishes faster than the susceptibility, so the charge diffusivity also vanishes. 

In the two models we consider, it thus appears that there is a close relation between the vanishing of the dc conductivity and the violation of the diffusivity bound. It would be interesting to prove that the static susceptibility can never vanish fast enough to spoil this.

To get a better handle on how higher-derivative couplings affect the bound on charge diffusion, it would be interesting to consider other models, such as non-linear electrodynamics \cite{Baggioli:2016oju}, including non-linear DBI setups \cite{Kiritsis:2016cpm}.\\

The higher derivative couplings we have considered do not affect the energy diffusion bound in the incoherent limit, including when an arbitrary potential $V(X)$ for the St\"uckelberg fields is included. A natural future direction would be to consider higher derivative couplings between the gravity and St\"uckelberg sector, responsible for momentum relaxation. A careful analysis of causality along the lines of \cite{Camanho:2014apa} will be required in this case. More recently, it was shown in \cite{Gu:2017ohj} that inhomogeneities could lead to a sign reversal of the bound. Understanding better the validity of the diffusion bounds featuring the butterfly velocity and the interplay with translation symmetry breaking is clearly an important issue.
\vskip 1cm

\section*{Acknowledgments}
We would like to thanks Mike Blake and Sean Hartnoll for interesting and useful comments on the manuscript.
MB would like to thank R. Nepomechie and the University of Miami for the warm hospitality during the completion of this work.

This work was supported in part by the Advanced ERC grant SM-grav, No 669288.
 BG is partially supported by the Marie Curie International Outgoing Fellowship nr 624054 within the 7th European Community Framework Programme FP7/2007-2013. The work of BG was partially performed at the Aspen Center for Physics, which is supported by National Science Foundation grant PHY-1066293. WJL is financially supported by  the Fundamental Research Funds for the Central Universities No. DUT 16 RC(3)097, NSFC Grants No. 11275208 as well as 11375026.\\

\appendix
\section{Equations of motion}
For the sake of completeness we show the equations of motion for the models considered in this short appendix.
\vskip.5cm
\paragraph{Model 1\\}
The $\mathcal{J}$ coupling is not affecting the background equations of motion. Therefore the latter coincide exactly with the EOMs for the linear St\"uckelbergs model presented in \cite{Andrade:2013gsa}. We omit them.
\vskip.5cm
\paragraph{Model 2\\}
In order to be coincise we define $Y\equiv F^2/4$.\\
The equations of motion for the model 2 defined in \ref{model2sec} generically read:
\begin{eqnarray}\label{geoms}
&\partial_\mu\left[\sqrt{-g}\,\mathcal{W}_{Y}(Y,X)\,F^{\mu\nu}\right]=0 \label{geoms1}\\
&\partial_\mu\left[\sqrt{-g}\,\mathcal{W}_{X}(Y,X)g^{\mu\nu}\partial_\nu\phi^I\right]=0 \label{geoms2} \\
&R_{\mu\nu}-\left[3+\frac{1}{2}R-\frac{1}{2}\,\mathcal{W}(Y,X)\right]g_{\mu\nu}\\ \nonumber
&=\frac{1}{2}\,\mathcal{W}_{Y}(Y,X)\,F_{\mu\sigma}{F_\nu}^\sigma+\frac{1}{2}\,\mathcal{W}_{X}(Y,X)\partial_\mu\phi^I\partial_\nu\phi^I \label{geoms3}.
\end{eqnarray}
Taking the ansatz \eqref{bansatz},  we obtain the equations of motion for $A_t$,  $B$, $C$ and $D$ as follows:
\begin{eqnarray}\label{seoms}
&\left[\mathcal{W}_{Y}\, \frac{C}{\sqrt{BD}}\, {A_t'}^2\right]'=0\\
&\left(6\,-\,\mathcal{W}\right)\,B\,D+\frac{C'^2\,D}{2\,C^2}+\frac{B'\,C'\,D}{B\,C}-\frac{2\,C''\,D}{C}
-\frac{1}{2}\,\mathcal{W}_{Y}\,{A_t'}^2=0,\\
&\left(\mathcal{W}-6\right)\,B\,D+\frac{C'\,D'}{C}+ \frac{1}{2}\,\frac{C'^2\,D}{C^2}+\frac{1}{2}\mathcal{W}_{Y}\,A_t'^2=0,\\
&\left(\mathcal{W}-6-\frac{k^2}{C}\,\mathcal{W}_{X}\right)\,B\,D+\frac{1}{2}\left(\frac{C'\,D'}{C}-\frac{B'\,C'\,D}{B\,C}\right)-\frac{1}{2}\frac{C'^2\,D}{C^2}+\frac{C''\,D}{C}+D''\\ \nonumber
&-\frac{1}{2}\left(\frac{B'\,D'}{B}+\frac{D'^2}{D}\right)-\frac{1}{2}\mathcal{W}_{Y}\,A_t'^2=0
\end{eqnarray}

\section{Derivation of the thermoelectric conductivities \label{app:DCcond}}

To compute the DC conductivities, we consider the following time-dependent perturbations around the background
\begin{eqnarray}\label{pansatz}
&\delta A_x=(\zeta A_t(r)-E) t+a_x(r),\nonumber\\
&\delta g_{tx}=-\zeta D(r) t+r^2h_{tx}(r),\nonumber\\
&\delta g_{rx}=r^2h_{rx}(r),\nonumber\\
&\delta\phi^I=\psi^x(r).
\end{eqnarray}
\paragraph{Model 1\\}
The equations of motion are given by
\begin{eqnarray}\label{A}
&\left[\left(1-\frac{\mathcal{J}\,k^2}{4\,r^2}\right)\left(D \,a_x'+\rho\, h_{tx}\right)\right]'=0\\  \label{A2}
&h_{rx}-\left(1-\frac{\mathcal{J}\,k^2}{4\,r^2}\right)\frac{\rho\,(\zeta A_t-E)}{k^2 \,\left(1+\frac{\mathcal{J}\,\rho^2}{4\,r^4}\right)\,D\,r^2}+\frac{r^2\,\zeta}{k^2 \left(1+\frac{\mathcal{J}\,\rho^2}{4\,r^4}\right)D}\left(\frac{D}{r^2}\right)'-\frac{{\psi^x}'}{k}=0\\ \label{A3}
&\left[r^2 \,\left(1+\frac{\mathcal{J}\,\rho^2}{4\,r^4}\right) D ({\psi^x}'-k\,h_{rx})\right]'-k\, \left(1+\frac{\mathcal{J}\,\rho^2}{4\,r^4}\right)\zeta=0\\ \label{A4}
&h_{tx}''+\frac{4}{r}\,h_{tx}'+\frac{k^2\left(1+\frac{\mathcal{J}\,\rho^2}{4\,r^4}\right)}{D\,r^2}h_{tx}+\frac{\rho}{r^{4}}\left(1-\frac{\mathcal{J}\,k^2}{4\,r^2}\right)a_{x}'=0.
\end{eqnarray}
We will adopt the strategy of \cite{Donos:2014cya} to express the currents $J^x$ and $Q^x$ in terms of horizon quantities.
From the Maxwell equation (\ref{A}), we define a conserved current along the radial direction in the bulk
\begin{eqnarray}\label{acurrent}
J\equiv-\left[\left(1-\frac{\mathcal{J}\,k^2}{4\,r^2}\right)\left(D\, a_x'+\rho\, h_{tx}\right)\right],
\end{eqnarray}
which one can check that it equals the U(1) current in the boundary theory
\begin{eqnarray}\label{ua1current}
<J^x>\equiv\frac{\delta S}{\delta A_x}\Big|_{r\rightarrow\infty}=-\lim_{r\rightarrow\infty}\sqrt{-g}\left[F^{rx}-\mathcal{J}\left(XF\right)^{[rx]}\right]
\end{eqnarray}
with the ansatz on fluctuations. Then we are going to construct a conserved current in the bulk which corresponds to the heat current on boundary $Q^x\equiv T^{t x}-\mu \,J^x$.
Finally, we find that the following quantity
\begin{eqnarray}\label{qcurrent}
Q=D^2\left(\frac{r^2h_{tx}}{D}\right)'-A_{t} J,
\end{eqnarray}
is constant along the radial direction, namely $\partial_r Q=0$. And one can further prove that the first term is related to the time-independent part of  the stress tensor $T_0^{tx}$ and the second term equals $\mu J^x$ as $r\rightarrow \infty$. Then $Q$ corresponds to the heat current $Q^x$ in the boundary theory.

The regular boundary conditions at the horizon can be chosen as follows
\begin{eqnarray}\label{rbc}
&&a_x\approx-\frac{E}{4\pi T}\,ln(r-r_h)+...\\
&&h_{tx}\approx D \,h_{rx}|_{r=r_h}-\frac{\zeta\, D}{4\,\pi \,T\,r_h^2}\,ln(r-r_h)+...
\end{eqnarray}
Then the electric and thermal currents can be expressed in terms of horizon quantities
\begin{eqnarray}\label{B}
&&J=\left[E\left(1-\frac{\mathcal{J}k^2}{4\,r^2}\right)\left(1+\left(1-\frac{\mathcal{J}k^2}{4\,r^2}\right)\frac{\rho^2}{k^2\left(1+\frac{\mathcal{J}\rho^2}{4\,r^4}\right) r^2}\right)+\zeta\left(1-\frac{\mathcal{J}k^2}{4\,r^2}\right)\frac{\rho \,D'(r)}{k^2 \left(1+\frac{\mathcal{J}\rho^2}{4\,r^4}\right)}\right]_{r=r_h},\nonumber \\
&&Q=\left[E \left(1-\frac{\mathcal{J}k^2}{4\,r^2}\right) \frac{\rho\, D'(r)}{k^2\left(1+\frac{\mathcal{J}\rho^2}{4\,r^4}\right)}+\zeta\,\frac{r^2\, D'(r)^2}{k^2 \left(1+\frac{\mathcal{J}\rho^2}{4\,r^4}\right)}\right]_{r=r_h}.
\end{eqnarray}
From these expressions the conductivities \eqref{DCmodel1} follow directly.\vskip.5cm
\paragraph{Model 2\\}
The equations of motion are given by
\begin{eqnarray}\label{feom}
&\left[\mathcal{W}_{Y}\left(\bar{Y},\bar{X}\right)D \,a_x'+\rho\, h_{tx}\right]'=0\\   \label{feom2}
&h_{rx}-\frac{\rho\,(\zeta A_t-E)}{k^2 \,\mathcal{W}_{X}\left(\bar{Y},\bar{X}\right)D\,r^2}+\frac{r^2\,\zeta}{k^2 \,\mathcal{W}_{X}\left(\bar{Y},\bar{X}\right) D}\left(\frac{D}{r^2}\right)'-\frac{{\psi^x}'}{k}=0\\ \label{feom3}
&\left[r^2 \,\mathcal{W}_{X}\left(\bar{Y},\bar{X}\right) D ({\psi^x}'-k\,h_{rx})\right]'-k\, \mathcal{W}_{X}\left(\bar{Y},\bar{X}\right)\zeta=0\\   \label{feom4}
&h_{tx}''+\frac{4}{r}h_{tx}'-\frac{k^2 \mathcal{W}_{X}\left(\bar{Y},\bar{X}\right)}{Dr^2}h_{tx}+\frac{\rho}{r^{4}}a_x'=0.
\end{eqnarray}
From the Maxwell equation (\ref{feom}), we define a conserved current along the radial direction in the bulk
\begin{eqnarray}\label{jcurrent}
J\equiv-[\mathcal{W}_{Y}\left(\bar{Y},\bar{X}\right)D \,a_x'+\rho\, h_{tx}],
\end{eqnarray}
which one can check that it equals the U(1) current in the boundary theory
\begin{eqnarray}\label{u1current}
<J^x>\equiv\frac{\delta S}{\delta A_x}\Big|_{r\rightarrow\infty}=-\lim_{r\rightarrow\infty}\sqrt{-g}\,\mathcal{W}_{Y}(\bar{Y},\bar{X})\,F^{r x}
\end{eqnarray}
with the ansatz on fluctuations. Then we are going to construct a conserved current in the bulk which corresponds to the heat current on boundary $Q^x\equiv T^{t x}-\mu J^x$.
Finally, we find that the following quantity
\begin{eqnarray}\label{qcurrent}
Q=D^2\left(\frac{r^2h_{tx}}{D}\right)'-A_{t} J,
\end{eqnarray}
is constant along the radial direction, namely $\partial_r Q=0$. And one can further prove that the first term is related to the time-independent part of  the stress tensor $T_0^{tx}$ and the second term equals $\mu J^x$ as $r\rightarrow \infty$. Then $Q$ corresponds to the heat current $Q^x$ in the boundary theory.

The regular boundary conditions at the horizon can be chosen as follows
\begin{eqnarray}\label{rbc}
&&a_x\approx-\frac{E}{4\pi T}\,ln(r-r_h)+...\\
&&h_{tx}\approx D \,h_{rx}|_{r=r_h}-\frac{\zeta\, D}{4\,\pi \,T\,r_h^2}\,ln(r-r_h)+...
\end{eqnarray}
Then the electric and thermal currents can be expressed in terms of horizon quantities
\begin{eqnarray}\label{currents}
&&J=\left[E\left(\mathcal{W}_{Y}\left(\bar{Y},\bar{X}\right)+\frac{\rho^2}{k^2\,\mathcal{W}_{X}\left(\bar{Y},\bar{X}\right)r^2}\right)+\zeta\frac{\rho \,D'(r)}{k^2 \,\mathcal{W}_{X}\left(\bar{Y},\bar{X}\right)}\right]_{r=r_h}, \\
&&Q=\left[E \,\frac{\rho \,D'(r)}{k^2 \,\mathcal{W}_{X}\left(\bar{Y},\bar{X}\right)}+\zeta\,\frac{r^2 \,D'(r)^2}{k^2 \,\mathcal{W}_{X}\left(\bar{Y},\bar{X}\right)}\right]_{r=r_h}.
\end{eqnarray}
Just taking the appropriate derivatives of the previous current we derive the conductivity matrix shown in \eqref{DCmodel2}.
\section{Background and thermoelectric conductivities for the specific models $2_U$ and $2_\mathcal{K}$}\label{appdetails}

In this appendix we will provide detailed formulae that give the background  and conductivities of the special 2 models.

\vskip .5cm

\paragraph{Model $2_U$\\}

\vskip 1cm

For this particular choice the background solution takes the form:
\begin{align}\label{c2bsolutions}
&A_t(r)=\rho\int_{r_h}^{r}\, \frac{1}{y^2\,U(k^2/y^2)}\,dy\,, \\
&D(r)=\frac{1}{2\,r}\int_{r_h}^{r}\, \left[6\,y^2-k^2-\frac{\rho^2}{2\,y^2\,U(k^2/y^2)}\right]\,dy\,.
\end{align}
and the Hawking temperature reads:
\begin{equation}\label{c2temperature}
T=\frac{1}{4\pi}\left[3r_h-\frac{1}{2}\frac{k^2}{r_h}-\frac{\rho^2}{4\,r_h^3U(k^2/r_h^2)}\right]
\end{equation}
The thermoelectric DC data are given by:
\begin{align}
&\sigma=U\left(X_h\right)+\frac{4\,\pi\,\rho^2}{k^2s\left(1-\frac{8\,\pi^2\,\rho^2\,U'(X_h)}{s^2\,U(X_h)^2 }\right)},   \\
&\alpha=\bar{\alpha}=\frac{4\,\pi\,\rho}{k^2\left(1-\frac{8\,\pi^2\,\rho^2\,U'(X_h)}{s^2\,U(X_h)^2 }\right)}, \\
&\bar{\kappa}=\frac{4\,\pi\, s\,T}{ k^2\left(1-\frac{8\,\pi^2\,\rho^2\,U'(X_h)}{s^2\,U(X_h)^2 }\right)}   \\
&\kappa=\frac{4\,\pi\, s\,T}{k^2\,\left(1-\frac{8\,\pi^2\,\rho^2\,U'(X_h)}{s^2\,U(X_h)^2 }\right)+\frac{4\,\pi\,\rho^2}{s\,U(X_h)}}.
\end{align}
where for convenience we defined $X_h=k^2/r_h^2$ and s is the entropy density $s=4\,\pi\,r_h^2$.\\

The chemical potential for the system can be defined as usual by:
\begin{equation}
\mu\,=\,A_t(\infty)\,-\,A_t(r_h)\,=\rho\,\int_{r_h}^{\infty}\, \frac{1}{y^2\,U(k^2/y^2)}\,dy
\end{equation}
\textit{i.e.} the leading value of the gauge field at the boundary once the regularity condition $A_t(r_h)=0$ is provided.

Since in the incoherent limit the radius of the horizon $r_h$ is just a function of the momentum dissipation strength $k$, it is straightforward to compute the susceptibility in that limit as:
\begin{equation}
\chi^{(inc)}\,=\,\frac{\partial \rho}{\partial \mu}\,=\,\left(\frac{\partial \mu}{\partial \rho}\right)^{-1}\,=\,\left(\int_{r_h}^{\infty}\, \frac{1}{y^2\,U(k^2/y^2)}\,dy\right)^{-1}
\end{equation}
which is the result presented in the main text.

\vskip .5cm

\paragraph{Model $2_\mathcal{K}$\\}

In this subsection we give more details about the solution for the $2_\mathcal{K}$ model.\\
Assuming the $U(X)$ function to be of the form:
\begin{equation}
U(X)\,=\,1\,+\,\mathcal{K}\,X
\end{equation}
the background solution for the gauge field is:
\begin{equation}\label{AtsolK}
A_t(r)\,=\,\frac{\rho  \left(ArcTan\left(\frac{r}{k\, \sqrt{\,\mathcal{K}}}\right)\,-\,ArcTan\left(\frac{r_h}{k  \,\sqrt{\mathcal{K}}}\right)\right)}{k\,  \sqrt{\mathcal{K}}}
\end{equation}
while the temperature and the electric DC conductivity are:
\begin{align}
&T\,=\,-\frac{\rho ^2}{16 \,\pi\,  r_h\,\left(k^2 \,\mathcal{K}+r_h^2\right)}-\frac{k ^2}{8 \,\pi  \,r_h}+\frac{3 \,r_h}{4\, \pi }\,\\
&\sigma_{DC}\,=\,1+\frac{\mathcal{K}\, k^2}{r_h^2}+\frac{\rho ^2}{k^2 \,r_h^2\, \left(1-\frac{\mathcal{K}\, \rho ^2}{2 \,\left(\mathcal{K}\, k^2+r_h^2\right)^2}\right)}\,
\end{align}
The other thermoelectric conductivities for the choice $U(X)=1+\mathcal{K}X$ can be directly extracted from the results above and for brevity we omit them.
Note that for $\mathcal{K}<0$ the solution for the gauge field in the $r$ coordinate becomes:
\begin{equation}\label{AtsolK}
A_t(r)\,=\,\frac{\rho  \left(ArcTanh\left(\frac{r_h}{k\, \sqrt{\,|\mathcal{K}|}}\right)\,-\,ArcTanh\left(\frac{r}{k  \,\sqrt{|\mathcal{K}|}}\right)\right)}{k\,  \sqrt{|\mathcal{K}|}}
\end{equation}
and it is clearly problematic. Indeed from the previous expression we see that:
\begin{equation}
\frac{r}{k  \,\sqrt{|\mathcal{K}|}}\,<\,1\,.
\end{equation}
which cannot be the case since the boundary is located at $r=\infty$.

In order to have a well defined solution we have to redefine the radial coordinate as follows:
\begin{equation}\label{gauge}
z\,=\,\sqrt{r^2\,-\,|\mathcal{K}|\,k^2}
\end{equation}
Of course all the physical quantities turn out to be independent of the radial coordinate choice and they are continuous with respect to the coupling $\mathcal{K}$. Note that $\mathcal{K}\geq -1/6$ for consistency.

In more detail, in this new radial coordinate we have that the functions appearing in the metric become:
\begin{equation}
C(z)\,=\,z^2\,+\,|\mathcal{K}|\,k^2\sp B(z)\,=\,\frac{z^2}{D(z) \left(k^2 \,|\mathcal{K}|+z^2\right)}
\end{equation}
The solution for gauge field is:
\begin{equation}
A_t(z)\,=\,\frac{\rho \, \log \left(\frac{z \left(k \sqrt{|\mathcal{K}|}+\sqrt{k^2 \,|\mathcal{K}|+z_h^2}\right)}{z_h\left(k\,
   \sqrt{|\mathcal{K}|}+\sqrt{k^2 \,|\mathcal{K}|+z^2}\right)}\right)}{k\, \sqrt{|\mathcal{K}|}}
\end{equation}
We can check that using this new radial coordinate there is no issue for any value in the range $-1/6\leq \mathcal{K}<0$.\\
The formula for the temperature gets modified into:
\begin{equation}
T\,=\,\frac{1}{4\,\pi}\,\frac{g_{tt}'(z_h)}{\sqrt{g_{tt}(z_h)\,g_{rr}(z_h)}}\,=\,\frac{1}{4\,\pi}\frac{D'(z_h)}{\sqrt{B(z_h)\,D(z_h)}}
\end{equation}
which gives:
\begin{equation}
T\,=\,-\frac{2 \,k^2 (1-6 \,|\mathcal{K}|) \,z_h^2+\rho ^2-12 \,z_h^4}{16 \,\pi  \,z_h^2 \sqrt{k^2 \,|\mathcal{K}|+z_h^2}}
\end{equation}
In the incoherent limit we now have:
\begin{align}
&z_h^{(inc)}\,=\,\frac{\sqrt{1-6\,|\mathcal{K}|}\,k}{\sqrt{6}}\,,\\
&\sigma_{DC}^{(inc)}\,=\,\frac{\left(z_h^{(inc)}\right)^2}{\left(z_h^{(inc)}\right)^2\,+\,|\mathcal{K}|\,k^2}\,=\,1\,-\,6\,|\mathcal{K}|\,,\\
&\chi^{(inc)}\,=\,\frac{k  \sqrt{|\mathcal{K}|}}{\log \left(\frac{k\,  \sqrt{|\mathcal{K}|}+\sqrt{k^2\, |\mathcal{K}|+\,\left(z_h^{(inc)}\right)^2}}{z_h^{(inc)}}\right)}\,=\,\frac{k\,  \sqrt{|\mathcal{K}|}}{\log \left(\frac{\sqrt{6\,|\mathcal{K}|}+1}{\sqrt{1-6 \,|\mathcal{K}|}}\right)}\,.
\end{align}
Considering also that the expression for the butterfly velocity in this new $z$ coordinate reads as follows:
\begin{equation}
v_B^2\,=\,\frac{\pi\,T}{\sqrt{z_h^2\,+\,|\mathcal{K}|\,k^2}}
\end{equation}
we arrive at the final expression for the charge diffusion appearing in the main text:
\begin{equation}
\frac{D_c\,T}{v_B^2}\Big|_{inc}\,=\,\frac{(1-6 \,|\mathcal{K}|) \log \left(\frac{ \sqrt{6\,|\mathcal{K}|}+1}{\sqrt{1-6 \,|\mathcal{K}|}}\right)}{\pi\, \sqrt{6\,|\mathcal{K}|}}
\end{equation}
which is valid for $\mathcal{K}<0$ and it joins continuosly with the expression for positive $\mathcal{K}$ as expected.

\section{Kelvin formula}\label{AppKelvin}
In this appendix we prove explicitely that the Kelvin formula:
\begin{equation}\label{Kformula}
\frac{\alpha}{\sigma}\Big|_{T=0}\,\equiv\,\lim_{T\rightarrow 0}\frac{\partial s}{\partial \rho}\Big|_T
\end{equation}
holds for all the models we considered.\\[0.2cm]
\paragraph{$\mathcal{J}$ model\\}
For this model the extremal horizon is located at:
\begin{equation}
r_0\,=\,\frac{\sqrt{2\,k^2\,+\,\mu^2}}{2\,\sqrt{3}}
\end{equation}
The Seebeck coefficient at zero temperature is:
\begin{equation}
\frac{\alpha}{\sigma}\Big|_{T=0}\,=\,\frac{2\,\pi\,\mu\,\sqrt{2\,k^2\,+\,\mu^2}}{\sqrt{3}\,(k^2\,+\,\mu^2)}
\end{equation}
Using the chain rule:
\begin{equation}
\frac{\partial s}{\partial \rho}\,=\,\frac{\partial s}{\partial \mu}\,\left(\frac{\partial \rho}{\partial \mu}\right)^{-1}
\end{equation}
and noticing that $\rho=\mu\,r_h$ and the finite temperature horizon is located at:
\begin{equation}
r_h\,=\,\frac{1}{6} \,\left(\sqrt{6 \,k^2+3\, \mu ^2+16\, \pi ^2 \,T^2}\,+\,4 \,\pi  \,T\right)
\end{equation}
it is straightforward to show that \eqref{Kformula} holds.\\[0.2cm]

\paragraph{$\mathcal{W}$ model\\}
In this model the Seebeck coefficient is generically given by:
\begin{equation}\label{check}
\frac{\alpha}{\sigma}\,=\,\frac{4 \,\pi \, \rho\,  r_h^2}{k^2\,r_h^2\,\mathcal{W}_Y\,
   \mathcal{W}_X\,+\,\rho ^2}\,\underbrace{=}_{Maxwell\,eq.}\,\frac{4 \,\pi \, r_h^2\, A_t'(r_h)}{\rho  \,A_t'(r_h)+k^2 \,\mathcal{W}_X}
\end{equation}
where its zero temperature value is just obtained replacing $r_h$ with the position of the extremal horizon $r_0$.\\
In order to compute the thermal derivative is convenient to use:
\begin{equation}
\frac{\partial s}{\partial \rho}\Big|_T\,=\,\frac{\partial s}{\partial r_h}\,\frac{\partial r_h}{\partial \rho}\Big|_T
\end{equation}
where the last term can be derived using the equation of state as follows:
\begin{equation}
dT\,=\,\frac{\partial T}{\partial r_h}\,d r_h\,+\,\frac{\partial T}{\partial \rho}\,d \rho\,=\,0
\end{equation}
Using the Maxwell equation the thermal derivative at fixed temperature becomes:
\begin{equation}
\frac{\partial s}{\partial \rho}\Big|_T\,=\,\frac{8 \,\pi\,  r_h^2 \,A_t'(r_h)}{\rho  \,A_t'(r_h)\,+\,2\, k^2 \,\mathcal{W}_X\,-\,2\, \Lambda \, r_h^2\,-\,r_h^2\, \mathcal{W}}
\end{equation}
Imposing the zero temperature limit:
\begin{equation}
\Lambda\,=\,\frac{1}{2} \left(-\mathcal{W}_Y\, A_t'(r_0)^2\,-\,\mathcal{W}\right)
\end{equation}
we obtain:
\begin{equation}
\lim_{T \rightarrow 0}\,\frac{\partial s}{\partial \rho}\Big|_{T}\,=\,\frac{4 \,\pi \, r_0^2\, A_t'(r_0)}{\rho  \,A_t'(r_0)+k^2 \,\mathcal{W}_X}
\end{equation}
which coincides with \eqref{check} at zero temperature.\\
In conclusion, also in the generic $\mathcal{W}$ model, the Kelvin formula holds.
\section{Null Energy Condition}\label{AppCons}
In this short appendix we summarize and give more details about the consistency analysis performed.\vskip.5cm
\paragraph{Model 1}
The consistency of this model has already been analyzed in \cite{Gouteraux:2016wxj} and constrains the coupling $\mathcal{J}$ to satisfy:
\begin{equation}
0\,\leq\,\mathcal{J}\,\leq 2/3
\end{equation}
We refer the reader to \cite{Gouteraux:2016wxj} for details.\vskip.5cm
\paragraph{Model 2\\}
Generically the NEC is given by
\begin{eqnarray}\label{NECa}
T_{\mu\nu}k^\mu k^\nu\ge0,
\end{eqnarray}
where $k^\mu$ is a null vector $k^\mu k_\mu=0$.  Recall that the stress tensor is
\begin{eqnarray}\label{NECa1}
T_{\mu\nu}=-g_{\mu\nu}\mathcal{W}(Y,X)+\mathcal{W}_Y(Y,X)\,F_{\mu\rho}F_{\nu\sigma}g^{\rho\sigma}+\mathcal{W}_X(Y,X)\partial_\mu\phi^I\partial_\nu\phi^I.
\end{eqnarray}
We then have
\begin{eqnarray}\label{NECb}
\mathcal{W}_Y(Y,X) F_{\mu\rho}F_{\nu\sigma}g^{\rho\sigma}k^\mu k^\nu+\mathcal{W}_X(Y,X)\partial_\mu\phi^I\partial_\nu\phi^Ik^\mu k^\nu\ge0
\end{eqnarray}
We can construct a complete basis for the null vectors space, which is given by
\begin{eqnarray}\label{basis}
&&k_{(1)}^\mu=(D(r)^{-1/2},D(r)^{1/2},0,0)\nonumber\\
&&k_{(2)}^\mu=(D(r)^{-1/2},0,1/r^2,0)\nonumber\\
&&k_{(3)}^\mu=(D(r)^{-1/2},0,0,1/r^2)
\end{eqnarray}
All in all, we derive the following constraint
\begin{eqnarray}\label{NECc}
\bar{X}\,\mathcal{W}_X(\bar{Y},\bar{X})-2\,\bar{Y}\,\mathcal{W}_Y(\bar{Y},\bar{X})\ge0
\end{eqnarray}
which is presented in the main text.\\[0.1cm]
Let us focus now on the benchmark model $\mathcal{W}(X,Y)=X+U(X)Y$ with $U(X)=1+ \mathcal{K}X$. It has already been proven in \cite{Gouteraux:2016wxj} that the coupling has to satisfy
\begin{equation}
-1/6\,\leq\,\mathcal{K}\,\leq\,1/6
\end{equation}
We checked the behaviour of various other quantities such as the heat capacity and the charge susceptibility in order to analyze the stability of the background solutions. As a result, we have not found stricter constraints than the ones already mentioned. The full analysis confirms the consistency range already obtained in \cite{Gouteraux:2016wxj}.

\section{Scaling dimensions of IR operators}\label{AppDim}
In this appendix we analyze the conformal dimension of the IR operators in the zero temperature limit. More concretely we study the transverse and longitudinal sectors of the linearized fluctuations around the AdS$_2\times R^2$ geometry.

The complete transverse and longitudinal sectors are defined by the following sets of (not independent) fluctuations:
\begin{align*}
&\text{transverse:   } \Big\{h_{ty},\,h_{xy},\,h_{uy},\,A_{y},\,\delta \phi^y\Big\}\\
&\text{longitudinal:  } \Big\{h_{tt},\,h_{xx},\,h_{yy},\,h_{tu},\,h_{uu},\,A_t,\,A_u,\,\delta \phi^x\Big\}
\end{align*}
where the momentum $q$ is taken for simplicity along the $x$ direction.\\
The correct way of proceeding would be to define gauge invariant independent variables but for simplicity we decide to work in gauge variant variables; in this way not all the fluctuations are independent and most of the equations read as constraints.

The AdS$_2\times R^2$ solution is defined by the following:
\begin{equation}
ds^2\,=\,-\,\frac{1}{u^2}\,dt^2\,+\,\frac{L_0^2}{u^2}\,du^2\,+\,dx^2\,+\,dy^2\,,\qquad A_t(u)\,=\,\frac{Q}{u}\,.
\end{equation}
The equations of motion fix the AdS$_2$ length $L_0$ and the IR charge $Q$ in terms of the cosmological constant $\Lambda$ and the momentum dissipation rate $k$.

In order to find the conformal dimensions of the IR operators we will perform a scaling ansatz for all the fields of the type:
\begin{equation}
\Psi_i(t,u,x)\,=\,\alpha_i\,u^{\Delta_i}\,e^{i\,(q\,x\,-\,\,\omega\,t)}
\end{equation}
where q and $\omega$ are the momentum and the frequency of the fluctuations. The power $\Delta_i$ is related to the conformal dimension of the IR operator dual to the bulk field $\Psi_i$ and $\alpha_i$ is just a normalization constant.\\
We will then solve the algebraic equations around the AdS$_2\times R^2$ background and extract the powers $\Delta_i$.

In order for the background to be stable the conformal dimensions of the IR operators, and more practically the solutions for $\Delta_i$, have to be real; this requirement could possibly constrain the possible values of the higher derivatives couplings.

For simplicity we focus just on the $\mathcal{J}$ and the $2_\mathcal{K}$ models and we omit most of the lengthy computations.

\vspace{0.2cm}
\paragraph{$\mathcal{J}$ model\\}
\vspace{0.2cm}
For the $\mathcal{J}$ model defined in sec. \ref{model1sec} the AdS$_2\times R^2$ solution is defined by:
\begin{equation}
L_0\,=\,\frac{2\,-\,Q^2}{k^2},\,\qquad \Lambda\,=\,\frac{(Q^2-4)\,k^2}{4\,(Q^2-2)}\,.
\end{equation}
In the following we will normalize the AdS$_2$ length to $1$ by fixing:
\begin{equation}
Q\,=\,\sqrt{2-k^2}
\end{equation}
This choice will force $k^2<2$.\\
In the transverse sector we adopt the radial gauge $h_{uy}=0$ and in order to find a solution we take the scaling ansatz:
\begin{equation}
A_y=\bar{a}_y\,u^{\Delta_T\,+\,3}\,,\,\,\,h_{ty}=\bar{h}_{ty}\,u^{\Delta_t}\,,\,\,\,h_{xy}=\omega\,\bar{h}_{xy}\,u^{\Delta_T\,+\,2}\,,\,\,\,\delta \phi^y=\omega\,\bar{\phi}^y\,u^{\Delta_T\,+\,4}\,.
\end{equation}
We can then solve for all the normalization constants (note that one of them is not physical and can be set to the identity) and consequently determine the power $\Delta_T$ which will fix the conformal dimensions of the IR operators in the transverse sector.\\
All in all we are left with the following equations for $\Delta_T$ (removing some modes which can be checked are pure gauge):
\begin{align*}
&4\,q^2\,+\left(-k^2+\Delta_T \,(\Delta_T+7)+12\right) \left(\left(k^2-2\right) \mathcal{J}-4\right)\,=\,0\,,\\[0.2cm]
&q^2 \left(-2 \,k ^2 ((3 \,\Delta_T (\Delta_T+5)+20) J+4)+16 (\Delta_T+2) (\Delta_T+3)+k^6 \left(-\mathcal{J}^2\right)+2 \,k^4\, \mathcal{J} (\mathcal{J}+3)\right)+\\
&+2
   (\Delta_T+1) (\Delta_T+2) (\Delta_T+3) (\Delta_T+4) \left(k^2 \,\mathcal{J}-4\right)+4 \,q^4 \left(k^2\, \mathcal{J}-2\right)\,=\,0\,.
\end{align*}
We can solve these equations with Mathematica and check that in the range:
\begin{equation}
0\,\leq\,\mathcal{J}\,\leq\,\frac{2}{3}
\end{equation}
all the roots are real.\\

We now proceed with the longitudinal sector. The independent fields are taken to be:
\begin{equation}
A_t=\bar{a}_t\,u^{\Delta_L\,+\,2}\,,\,\,\,h_{tt}=\bar{h}_{tt}\,u^{\Delta_L}\,,\,\,\,h_{xx}=\bar{h}_{xx}\,u^{\Delta_L\,+\,2}\,,\,\,\,h_{yy}=\bar{h}_{yy}\,u^{\Delta_L\,+\,2}\,,\,\,\,\delta \phi^y\,=\,\frac{\bar{\phi}^y}{k}\,u^{\Delta_L\,+\,2}\,.
\end{equation}
Following a similar procedure we obtain the equations:
\begin{align*}
&k^4 \left(-4\, q^2 ((\Delta_L+1) (\Delta_L+2) \mathcal{J}+1)+\Delta_L (\Delta_L+1) (\Delta_L+2) (\Delta_L+3) \mathcal{J}+2 \,\mathcal{J}\, q^4\right)+\\
&+2 \left(-2 (\Delta_L+1)
   (\Delta_L+2) q^2+\Delta_L (\Delta_L+1) (\Delta_L+2) (\Delta_L+3)+q^4\right) \\ &\Big((\Delta_L+1) (\Delta_L+2) (\mathcal{J}+2)-2 \,q^2\big)-k^2
   \big(q^4 (3 (\Delta_L+1) (\Delta_L+2) \mathcal{J}+8)\\&-(\Delta_L+1) (\Delta_L+2) q^2 ((3 \Delta_L (\Delta_L+3)+10) \mathcal{J}+12)+\Delta_L (\Delta_L+1)
   (\Delta_L+2) (\Delta_L+3) \\&((\Delta_L (\Delta_L+3)+4) \mathcal{J}+4)-\mathcal{J}\, q^6\Big)+k^6 \,\mathcal{J}\, q^2\,=\,0\,.
\end{align*}
Again it is possible to prove that once we restrict:
\begin{equation}
0\,\leq\,\mathcal{J}\,\leq\,\frac{2}{3}
\end{equation}
all the roots are real.\\
\vspace{0.2cm}
\paragraph{$\mathcal{K}$ model\\}
\vspace{0.2cm}
We can run the same arguments as before for the $2_{\mathcal{K}}$ model.\\
The AdS$_2\times R^2$ solution is now defined by:
\begin{equation}
L_0\,=\,\frac{4\,-\,2\,Q^2\,-\,\mathcal{K}\,Q^2\,k^2}{2\,k^2},\,\qquad \Lambda\,=\,\frac{(Q^2\,-\,4)\,k^2}{2\,(4\,-\,2\,Q^2\,-\,\mathcal{K}\,Q^2\,k^2)}\,.
\end{equation}
We can again normalize the AdS$_2$ length to $1$ by fixing:
\begin{equation}
Q\,=\,\sqrt{\frac{2\,(2\,-\,k^2)}{2\,+\,\mathcal{K}\,k^2}}
\end{equation}
This choice will force $\frac{2\,-\,k^2}{2\,+\,\mathcal{K}\,k^2}>0$.\\
For the transverse sector we are left with the following equations:
\begin{align*}
&k^2 \left(\mathcal{K} \left(\Delta_T (\Delta_T+7)-q^2+14\right)-2\right)+2 \left(\Delta_T (\Delta_T+7)-q^2+12\right)-2 \,k^4\, \mathcal{K}\,=\,0\,,\\[0.2cm]
& 2 \left(\Delta_T^4+10 \Delta_T^3+35 \Delta_T^2-2 (\Delta_T+2) (\Delta_T+3) q^2+50 \Delta_T+q^4+24\right)+\\&+k^2 \left(\mathcal{K}
   \left(\Delta_T (\Delta_T+3)-q^2+2\right) \left(\Delta_T (\Delta_T+7)-q^2+12\right)+2 q^2\right)+2 \,k^4\, \mathcal{K}\, q^2\,=\,0\,.
\end{align*}
and again if:
\begin{equation}
-1/6\,\leq\,\mathcal{K}\,\leq\,1/6
\end{equation}
all the roots are real.\\
In the longitudinal sector it is convenient to perform the following redefinition:
\begin{equation}
\Delta_L\,=\,\frac{-3+\sqrt{\tilde{\Delta}_L}}{2}
\end{equation}
In that way the equation for $\tilde{\Delta}_L$ becomes of the cubic form:
\begin{equation}
a\,\tilde{\Delta}_L^3\,+\,b\,\tilde{\Delta}_L^2\,+\,c\,\tilde{\Delta}_L\,+\,d\,=\,0
\end{equation}
where:
\begin{align*}
a\,=&\,\left(k^2 \,\mathcal{K}+1\right)^2 \left(k^2\, \mathcal{K}+2\right)^2 \left(\left(k ^2-1\right) \mathcal{K}+1\right)\,,\\[0.2cm]
b\,=&\,(-\,k^2 \mathcal{K}-1) (k^2\,\mathcal{K}+2)(8 \,k^2+\mathcal{K} (8 \,k^8 \,\mathcal{K}^2+k^6\, \mathcal{K}
   (\mathcal{K} (8 q^2-5)+24)\\&+k^4 (-4 (\mathcal{K}-10) \mathcal{K}\, q^2-3 (\mathcal{K}-4) \mathcal{K}+24)\\&+k^2
   (-20 (\mathcal{K}-3) q^2-25 \,\mathcal{K}+39)-24 q^2-22)+24 q^2+22)\,,\\[0.2cm]
   c\,=&\,(k^2 \mathcal{K}+1) (16 k^{10} \mathcal{K}^4 (2 q^2+5)+k ^8 \mathcal{K}^3 (32 \mathcal{K} q^4+16 (\mathcal{K}+14)
   q^2-141 \mathcal{K}+400)\\&+k^6 \mathcal{K}^2 (\mathcal{K} (-16 (\mathcal{K}-14) q^4-8 (\mathcal{K}-14) q^2+61 \mathcal{K}-526)+544
   q^2+720)\\&+k ^4 \mathcal{K} (145 \mathcal{K}^2-112 (\mathcal{K}-5) \mathcal{K} q^4+8 ((47-7 \mathcal{K}) \mathcal{K}+68) q^2-553
   \mathcal{K}+560)\\&+4 k^2 (16 (9-4 \mathcal{K}) \mathcal{K} q^4+8 ((17-6 \mathcal{K}) \mathcal{K}+6) q^2+\mathcal{K} (2 \mathcal{K}-23))\\&-4
   \mathcal{K} (48 q^4+56 q^2+19)+192 q^4+4 (40 k^2+56 q^2+19))\,,\\[0.2cm]
   d\,=&\,(k^2 \mathcal{K}+1) (k^6 (-\mathcal{K}^4) (4 q^2+9) (8 k ^4-15 k ^2+16 (k
   ^2-1) q^4+4 (9 k ^2-8) q^2+7)\\&-k ^4 \mathcal{K}^3 (4 q^2+9) (40 k ^4-58 k ^2+16 (6
   k^2-5) q^4+8 (8 k ^4+4 k ^2-5) q^2+19)\\&+k ^2 \mathcal{K}^2 (-9 (72 k ^4-67 k
   ^2+8)+64 (8-13 k ^2) q^6-16 (64 k ^4+67 k ^2-48) q^4\\&-4 (64 k ^6+328 k ^4-5 k
   ^2-32) q^2)+4 \mathcal{K}(-126 k ^4+45 k ^2+64 (1-3 k ^2) q^6\\&-16 (20 k ^4+9 k ^2-3)
   q^4-4 (32 k ^6+50 k ^4+25 k ^2-11) q^2+9)\\&-4 (4 k ^2+4 q^2+1) (8 q^2 (2 k ^2+2
   q^2+1)+9))\,.
\end{align*}
In principle we have to show that all the roots of such a cubic equation in terms of the new variable $\tilde{\Delta}_L$ are real and positive. In order to do so it is better to recast the equation in the form:
\begin{equation}
t^3\,+\,p\,t\,+\,q\,=\,0\,.
\end{equation}
using the change of variable:
\begin{equation}
\tilde{\Delta}_L\,\longrightarrow\,t\,-\,\frac{b}{3\,a}
\end{equation}
and define the discriminant:
\begin{equation}
\mathcal{D}\,=\,\frac{q^2}{4}\,+\,\frac{p^3}{27}
\end{equation}
If the discriminat is negative:
\begin{equation}
\mathcal{D}\,<\,0
\end{equation}
the cubic equation has 3 real roots. Additionally if:
\begin{equation}
a\,b\,<\,0\,.
\end{equation}
all the roots are positive.\\
Because of the complexity of the expressions we have not been able to prove analytically the previous statements. Nevertheless we have performed several numerical checks and plots in order to assess their validity. We have found that in the range:
\begin{equation}
-1/6\,\leq\,\mathcal{K}\,\leq\,1/6
\end{equation}
no imaginary root appears.\\[0.2cm]
\paragraph{Final outcome\\}\strut\\[0.2cm]
The analysis of the conformal dimensions of the IR operators in the transverse and longitudinal sectors does not constraint further the range of validity of our higher derivative theories.
\bibliographystyle{JHEP}
\bibliography{HigherDerivativeAxions}
\end{document}